\documentclass[pdflatex,sn-basic]{sn-jnl}

\jyear{2024}%

\theoremstyle{thmstyleone}%
\theoremstyle{thmstyletwo}%

\theoremstyle{thmstylethree}%

\begin{document}

\title[Atmospheric escape from exoplanets]{Atmospheric escape from exoplanets: recent observations and theoretical models}

\author*[1]{\fnm{Gopal} \sur{Hazra}}\email{hazra@iitk.ac.in}

\affil*[1]{\orgdiv{Department of Physics}, \orgname{Indian Institute of Technology Kanpur}, \orgaddress{\street{Kalyanpur}, \city{Kanpur}, \postcode{208016}, \state{Uttar Pradesh}, \country{India}}}


\abstract{The review aims to give an overview of atmospheric escape processes from exoplanets. I briefly discuss the physics of various escape processes responsible for atmospheric escape across different types of exoplanets. Transmission spectroscopy is one of the major workhorses to observe the escaping atmosphere from exoplanets. I discuss recent observations that established the fact that atmospheric escape is very common in exoplanets, especially during the early phase of their evolution {when stellar high-energy radiation (X-ray and extreme ultraviolet, hence XUV) is strong}. There are many theoretical efforts/models to understand atmospheric escape processes. Stellar radiation is one of the major drivers of atmospheric escape, but other stellar environments (e.g., stellar flares, stellar winds, stellar coronal mass ejections, and stellar magnetic field) also have control over how the escape process will be affected for a given property of exoplanet, as the planetary properties (e.g., gravity, thermal energy, magnetic field) plays an important role for atmospheric escape. I discuss all governing factors for the atmospheric escape process and corresponding theoretical models in detail. I also discuss how atmospheric escape plays a crucial role in the overall atmospheric evolution of exoplanets and can lead us to understand some features in recently observed exoplanet demographics.}




\maketitle

\section{Introduction}\label{sec1}
The atmosphere of any planet is nothing but a collection of a large number of particles of different species bound to the planet. Mostly these particles consist of molecules, neutral atoms, and ions. Depending on the surface temperature of the planet and electromagnetic radiation from their host star, these particles get ionized. Hence, the study of the atmosphere means the study of collective phenomena these large groups of particles execute under different physical conditions (e.g., pressure, temperature, radiation, gravity, and magnetic field). To develop a theory of these {dynamical} systems where we want to understand the time evolution, Boltzmann's theory would be very much applicable to study their collective behavior if there is no external force working on them and they are non-interacting. For example, the dynamic behavior of the atmosphere on our planet Earth could be described by the Boltzmann theorem assuming the atmosphere consists of N non-interacting classical particles, which shows that the atmospheric particles relax to equilibrium if no external forces act on them giving rise to the Maxwellian velocity distribution \citep{Maxwell1860}. However, note that this is a very idealistic case and planetary atmospheres go through many external forces (e.g., radiation, ram pressure associated with the stellar wind and coronal mass ejections, thermal heating from the planetary core, and also both stellar and planetary magnetic field) from short time scale (days to years) to long time scales (Myrs) when it orbits around their host stars. Also, the constituent atoms of the atmosphere get ionized initiating a chemical reaction and the particle can no longer be assumed to be non-interacting. These combined effects make the study of planetary atmospheres far more interesting and complicated. 

Apparently, if someone closely observes the atmosphere around us on Earth, they will find that the atmosphere is very much neutral in the lower part where the temperature is on average close to the 288 K, unless we move to the upper layer in the atmosphere where radiation from the Sun is being absorbed. Hence, not in an unreasonable approximation, globally for the Earth, we could find that the whole atmosphere is in thermal equilibrium following the Maxwellian velocity distribution. Some of the particles that are in the tail of this Maxwellian distribution have more energy than the mean velocity of atmospheric particles and they escape to space overcoming Earth's gravity of influence. This is a special type of escape known as the Jeans escape where the escape happened on a particle-to-particle basis. In the case of the Earth, mostly low-mass atoms (e.g., hydrogen and helium) are able to escape the atmosphere. On present-day conditions, the observed atmospheric mass loss rates are 0.5 $\times$ 10$^3$ g~s$^{-1}$ for Venus, 1.4 $\times$ 10$^3$ g~s$^{-1}$ for the Earth, and between 0.7 $\times$ 10$^3$ g~s$^{-1}$ and 2.1 $\times$ 10$^3$ g~s$^{-1}$ for Mars respectively \citep{Gunell2018}. For Mars, the mass loss estimation was within some uncertainty as the atmospheric escape process for Mars gets disturbed by the dissociative recombination of molecular oxygen ions. For Pluto, the atmospheric mass loss ranged from as low as 0.69 $\times$ 10$^3$ g~s$^{-1}$ to 930.26 $\times$ 10$^3$ g~s$^{-1}$ \citep{Brain2016}. Different ranges of mass-loss for solar system planets that have different physical conditions and different distances from the Sun make it evident that planets go through different escape processes via which they lose their atmosphere. With the discovery of more than 5000 exoplanets, we have a much larger sample for which the atmospheric escape process might happen. So far, we have already observed many exoplanets that are going through active mass loss \citep[e.g.,][]{Vidal-Madjar2003, Vidal-Madjar04, Ehrenreich2015, Spake2018, Cubillos2020}. By studying atmospheric escape processes from exoplanets, we can put tighter constraints on them as different exoplanet environments give us extreme conditions compared to solar system planets.  

Largely, atmospheric escape processes are classified into two categories in literature: Thermal escape and non-thermal escape. The thermal escape is driven by the thermal processes which are controlled by stellar radiation mostly X-ray, Extreme ultraviolet flux, and infrared \citep[e.g.,][]{Chamberlain1963, Chamberlain1971, Watson1981, Murray-Clay2009}. Sometimes thermal escape is also driven by core-powered heat from the planets \citep{Ginzburg2018}. Hydrodynamic escape and Jeans escape belong to the category of thermal escape from the planet. The non-thermal escape is driven by the complex interaction of stellar wind with the planetary atmosphere including its magnetic counterpart. The non-thermal escape is classified into photochemical loss due to photo-chemical reaction \citep{Shematovich1994, Gronoff2007, Johnstone2018}, ion loss \citep{Garcia-sage2017}, ionospheric outflow \citep{Bouhram2004}, and instability-driven escape such as Kelvin-Helmholtz instability, Rayleigh-Taylor instabilities, and ion-ion instabilities \citep{Dubinin2011}.  

Atmospheric escape from the solar system planet has been observed from a long time ago \citep[e.g.,][]{Damon1957, Mayne1957, Gronoff2008, Chaffin2014}. By observing airglow emission, one can infer the density of escaping species and exospheric temperature leading to an estimate of the atmospheric mass loss rate \citep{Chaffin2014}. Also, in some cases, in situ spectrometry of neutrals and ions gives measurements of neutral and ion escapes \citep{Cui2008}. The airglow emission and in situ measurements for exoplanets are not possible for obvious reasons and we have different techniques to find atmospheric escape from exoplanets. To measure atmospheric escape from the exoplanets, transmission spectroscopy has been widely used. In transmission spectroscopy, planetary transits across the stellar disk are observed to compute the amount of stellar spectra being absorbed by the planetary atmosphere. The first observational evidence of escaping atmosphere was reported from the hot Jupiter HD209458b \citep{Vidal-Madjar2003}. Hydrogen is the more abundant and prone to escape from any planet and \citet{Vidal-Madjar2003} correctly searched for the hydrogen Lyman-$\alpha$ emission line using Space Telescope Imaging Spectrograph (STIS) onboard the Hubble Space Telescope (HST). They reported $15 \pm 4 \%$ of planetary atmospheric absorption of atomic hydrogen. The only explanation for this large amount of observed absorption is that the atmosphere is extended beyond the Roche limit. Hence it is attributed in terms of escaping hydrogen atoms. The first successful observation of escaping hydrogen atoms motivates further study of escaping hydrogen atoms and other escaping atmospheric species including heavy metal elements. Along with hydrogen \citep[e.g.,][]{LecavelierdesEtangs2010, Ehrenreich2011, Barnes16} and helium {\citep[e.g.,][]{Spake2018, Nortmann2018, Zhang2022b, Allart2023, Guilluy2023, Krishnamurthy24, McCreery2025}}, it also has been observed that oxygen, carbon, silicon, sodium, magnesium, ionized iron are escaping \citep[e.g.,][]{Vidal-Madjar04, Sing2019}, which make atmospheric escape as a significant phenomenon that is going on in exoplanets.         

\begin{figure}[htbp!]
\centering
\includegraphics[width=0.42\textwidth]{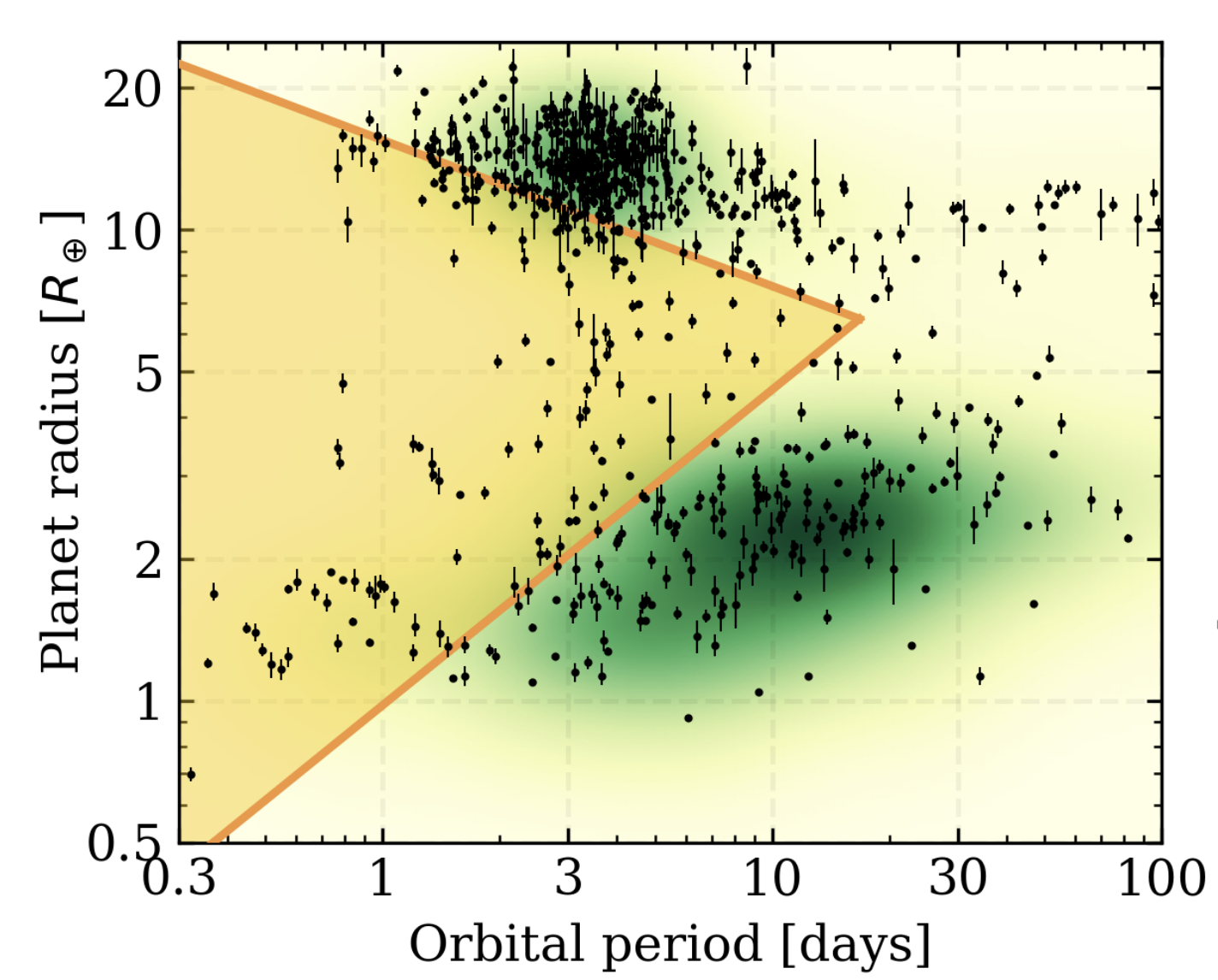}
\includegraphics[width=0.5\textwidth]{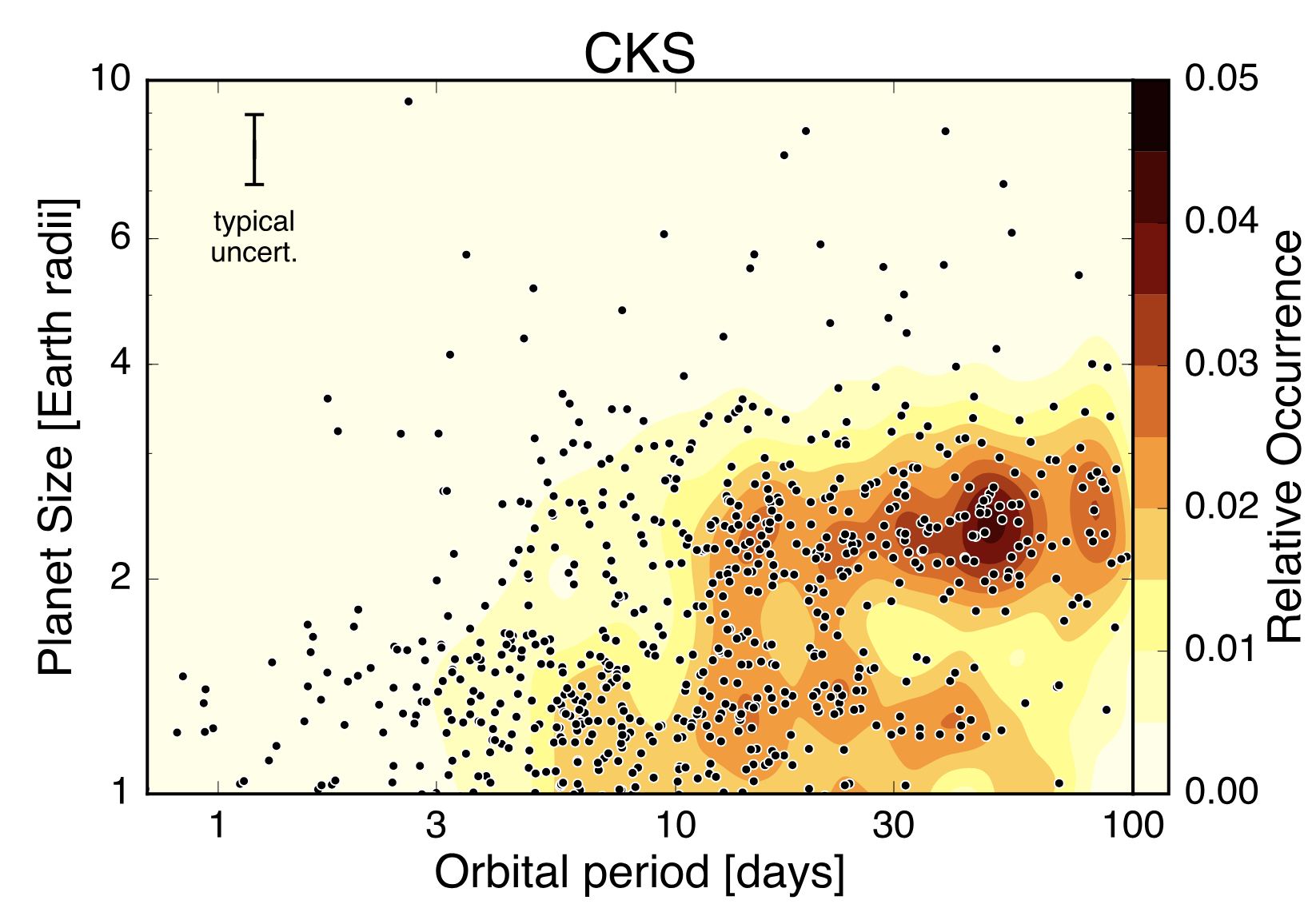}
\caption{{Left panel: the Neptunian desert in the observed exoplanet demographics (in the parameter space of planet size and orbital period). The yellow-shaded region under the triangular boundary shows the Neptunian desert area. The triangular desert boundary is defined by \citet{Mazeh2016}. The figure is taken from \citet{Cloutier2024} who modified the original version from \citet{Osborn2023}. The filled contours show the number density of confirmed exoplanets and black markers show the observed planets from NASA exoplanet archive. Right panel: radius valley observed in the small exoplanets from  California-Kepler Survey (CKS) from \citet{Fulton2017}.}}\label{fig:neptune_radius_valley}
\end{figure}

The study of atmospheric escape is not only important to understand the physics of the escape processes, but also it is very important for the evolution of the planetary atmosphere. A strong mass loss due to atmospheric escape will lead to no atmosphere on the planet over time. {When a planet evolves with time, it receives less XUV radiation from the host star, as the stellar activity declines with age due to spin down \citep{Tu2015}. During the early phase of planetary evolution, the XUV radiation remains intense leading to enhanced mass-loss from the planets \citep{Lopez2013, Owen_wu2016, Kubyshkina2018b, Kubyshkina2022}}. Current observed exoplanet demographics show two interesting features: A hot Neptunian desert \citep{Mazeh2016} and the Radius valley \citep{Fulton2017}. {The Neptunian desert as shown in the left panel of figure~\ref{fig:neptune_radius_valley} is a place in the radius-period plot of exoplanets, where the number of hot Neptunes of radius between 2-9 R$_\oplus$ and masses 10 - 250 M$_\oplus$ with an orbital period of less than 4 days, found to be very small.}
For the sub-Neptunian planets, there is a gap in the number of planets found within the radius of 1.9-2 R$_\oplus$ which is known as Radius Valley \citep{Fulton2017}. {The Radius Valley is shown in the right panel of figure~\ref{fig:neptune_radius_valley}}. Both the Neptunian desert and the Radius valley are explained by mass loss over time due to atmospheric escape \citep{Ionov2018, Rogers2021, Vissapragada2022}. However, sometimes planet formation scenarios are also invoked as a reason for their existence \citep{Venturini2020}. Overall, atmospheric escape is one of the key drivers for the evolution of the planetary atmosphere including its chemical compositions. 

The plan for the review is as follows. In the next section, we explain the physical processes that are responsible for atmospheric escape. The observational aspects of atmospheric escape for different exoplanets (e.g., gas giants, rocky exoplanets) are discussed in section~\ref{sec:observations}. In section~\ref{sec:numerical_model}, I explain various theoretical models that incorporate different atmospheric escape processes and their results including comparison with the observations. The evolution of the planetary atmosphere and the role of stellar environments in the evolution process is discussed in section~\ref{sec:evolution}. Finally, I will summarise the review with future outlooks in section~\ref{sec:conclusion}.

\section{Physical processes of atmospheric escape}\label{sec2}
In this section, I will review all the processes that have been suggested in the literature for atmospheric escape. I aim to explain the main parameters for each escape process and I will discuss whether it is possible to relate each escape process with some observable. As we discussed before, the atmospheric escape processes are largely divided into two main procedures based on their driving mechanisms: thermal escape and non-thermal escape. We discuss below both of the escape mechanisms in detail.

\subsection{Thermal escape}
Thermal escape is one of the important processes which is responsible for atmospheric loss from exoplanets. The X-ray and extreme ultraviolet radiation (hence XUV) from the host star mostly control the thermal escape from the exoplanets \citep{Garcia-Munoz2007}. Depending on the intensity of the radiation, thermal escape has two limits - Jeans escape and hydrodynamics escape. 
\subsubsection{Jeans escape}\label{sec:jeans}
\begin{figure}[htbp!]
\centering
\includegraphics[width=1.0\textwidth]{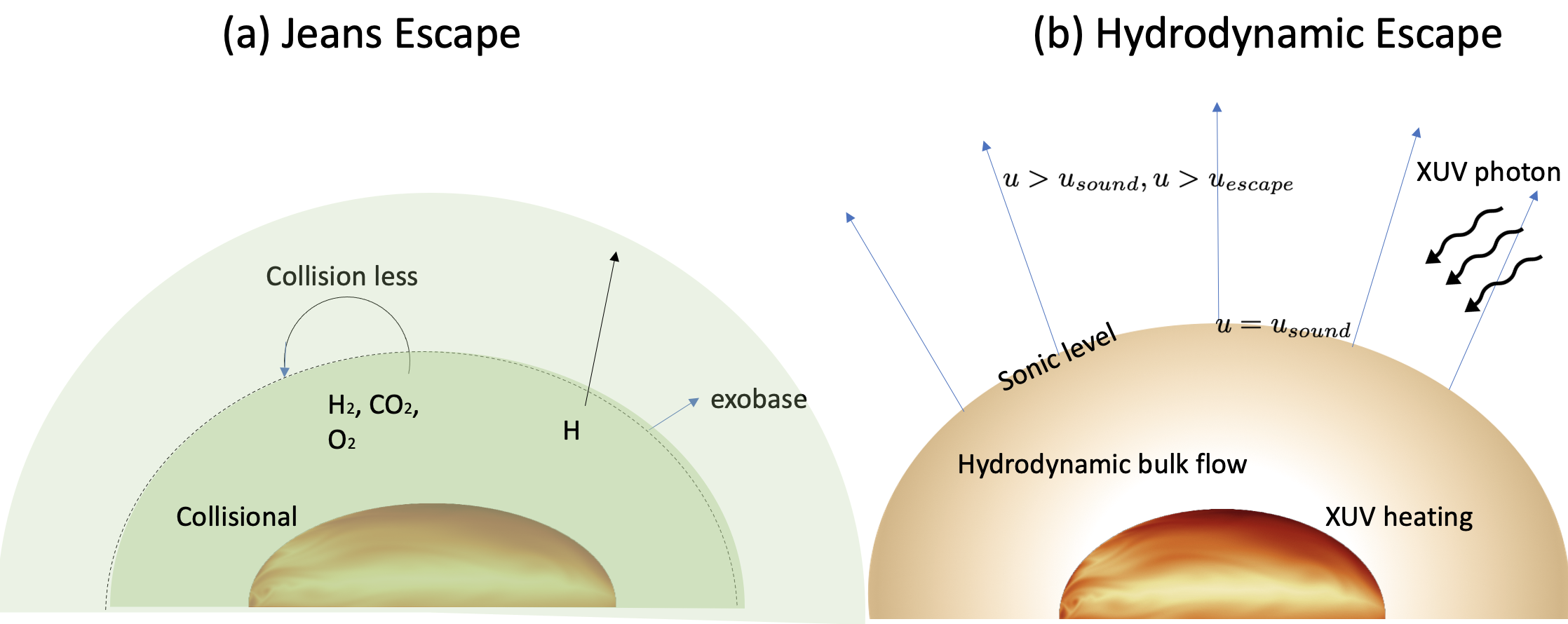}
\caption{Schematic diagram of different types of thermal escape. (a) Jeans escape - atmospheric particles escape from a collisional atmosphere to a collisionless regime. (b) Hydrodynamic escape - Collsional outflow due to stellar radiation heating mostly by X-ray and extreme ultraviolet part of the spectra }\label{fig:escape}
\end{figure}

If the upper atmosphere of the exoplanets is in local thermodynamic equilibrium or close to it, then their distribution function can be approximated by a Maxwellian distribution, which is given below following \citet{Gronoff2020}:
\begin{equation}
f(\vec{x},\vec{v}) = N\left(\frac{m}{2\pi kT}\right)^{3/2} e^{-\frac{mv^2}{2kT}} = N\left(\frac{1}{u_i\sqrt{\pi}}\right)^3 e^{-v^2/u_i^2},
\end{equation}
where $\vec{v}$ and m are the average velocity and mean mass of atmospheric particles respectively. T is the temperature of the upper atmosphere and $u_i$ = $\sqrt{\frac{2kT}{m_i}}$ is the velocity of the individual species. If some of the atmospheric species achieve escape velocity $v_{\rm esc} = \sqrt{\frac{2GM}{r}}$, they escape to space. The particle in this type of escape process moves from the collisional regime to the collisionless regime as shown in figure~\ref{fig:escape}(a). The boundary between these two regimes is called exobase where the mean free path of the escaping species is greater than the scale height. Also, it has been inherently assumed that the escaping species would not impact another molecule while escaping. 

An important parameter that is introduced to account for the Jeans escape in planetary atmospheres is called Jeans parameter $\lambda_{esc}$. This parameter is defined as the ratio of gravitational energy to the thermal energy-
\begin{equation}\label{eq:jeans}
\lambda_{esc} = \frac{GMm_i/r}{kT} = \frac{v_{\rm esc}^2}{ u_i ^2} 
\end{equation}
For the $\lambda_{esc}  > 2.5$, the Jeans escape becomes significant. To calculate, the escape flux of escaping atmospheric particles, we approximate the distribution to be Maxwellian at the exobase and integrate the vertical flux, $u_icos\theta \times f_i$ over the planet surface \citep{Gronoff2020}, which gives 
\begin{equation}
\phi_i(\rm escape) = N_i\left(\frac{u_i}{2\sqrt{\pi}}\right)(1 + \lambda_{esc})e^{-\lambda_{esc}}.
\end{equation}
However, as the particles that are faster than the escape velocity already left to space, a non-Maxwellian correction must be applied to consider this depletion of the high energy tail of Maxwellian distribution \citep{Chamberlain1971}. To capture the Jeans escape realistically one should solve the Boltzmann equation and there were few efforts to solve this using the Direct Monte Carlo simulation method \citep[][]{Volkov2011}. 

\subsubsection{Hydrodynamic escape}\label{hydroescape}
Hydrodynamic escape is a more dramatic version of Jeans' escape. If the internal energy of the individual particle of the atmosphere approaches the kinetic energy for escape, then the gas will escape as a continuous fluid. In hydrodynamic escape, the bulk of the collisional atmosphere leaves the planet as planetary outflow which is similar to the Parker wind (see figure~\ref{fig:escape}(b)). This type of atmospheric escape happens when the planet is irradiated with intense stellar radiation. The fundamental distinction between the Jeans escape and hydrodynamic escape is that atmospheric particles leave from the collisional domain to the collisionless domain by particle to particle but for hydrodynamic escape, the whole bulk collisional atmosphere leaves the planet as hydrodynamic fluid flow. 

The hydrodynamic escape is modeled using the same equation as the thermal Parker wind with an added external heating source. In Parker wind, the flow gets accelerated by gas pressure from subsonic to supersonic via a critical sonic point but in hydrodynamic escape, the external heating from stellar UV irradiation adds a major source of fluid acceleration. The hydrodynamic outflow originates from the sub-stellar point on the planet and mass-loss occurs in the form of a steady transonic wind. The basic equations that govern the hydrodynamic escape are the mass continuity, momentum conservation, and energy equation in the steady state:

\begin{eqnarray}
\frac{\partial}{\partial r}(r^2 \rho u) = 0  \label{eq:cont}\\
u\frac{\partial u}{\partial r} = -\frac{1}{\rho}\frac{\partial P_T}{\partial r} - \frac{GM_{\rm pl}}{r^2} + \frac{3GM_{\star}r}{a^3} \label{eq:mom}\\
\rho u \frac{\partial}{\partial r}\left[\frac{k_B T}{(\gamma -1)m}\right] = \frac{k_BT}{m}u\frac{\partial \rho}{\partial r} + \cal{H} - \cal{C}, \label{eq:energy}
\end{eqnarray}
where the notations are as usual. $\cal{H}$ and $\cal{C}$ represent the heating and cooling rates of the atmosphere, which are modeled separately.  In the momentum equation, in addition to gravity force, a tidal gravity term is added in the last part. The tidal gravity term includes a summation of centrifugal force and differential stellar gravity along the ray joining the planet to the star. In the energy equation, the heating due to stellar radiation is implemented. For simplicity, it is a general practice to assume the stellar XUV flux is concentrated at one photon energy $h\nu_0$ = 20 eV and calculate the corresponding heating rate as $\Gamma = \epsilon F_{\rm xuv} e^{-\tau}\sigma_{\nu_0}n_0$ \citep{Garcia-Munoz2007, Murray-Clay2009, Allan2019}. In this calculation, $n_0$ is the number density of neutral H atoms, $\epsilon = (h\nu_0 - 13.6 eV)/h \nu_0$ is the efficiency representing the fraction of photon energy deposited as heat, and $\tau$ is the optical depth. Note that \citet{Kubyshkina2020} and their group assume that the X-ray energy of stellar spectra is concentrated on one wavelength and extreme ultraviolet is distributed over another wavelength. {\citet{Hazra2020} implemented the heating term using full spectral energy distribution of XUV and concluded that their non-monochromatic approximation gives very similar outcomes as the monochromatic approximation, if an appropriate mean energy (i.e., the XUV radiation is only concentrated on a single wavelength of photon energy 20 eV) is adopted .} For the gas giants' atmosphere, different treatments of the stellar radiation do not affect the mass loss rate significantly \citep{Hazra2020} but for the small rocky exoplanets, full spectral energy distribution is necessary because they consist of many atmospheric species and photochemical reaction becomes important in all wavelength ranges \citep{Johnstone2018, Johnstone2020}.

Radiation cooling plays an important role in the cooling term $\cal{C}$ of equation~\ref{eq:energy}. For the gas giant atmosphere, the main contribution comes from Ly-$\alpha$ radiation. {However, the $H^+_3$ cooling \citep{Yelle2004, Koskinen2007} and metal cooling \citep{Huang_2023} are also important, even they become dominant in some regions of the upper atmosphere.} For the rocky exoplanets, the infrared cooling by CO$_2$, H$_2$O, NO and O becomes important \citep{Johnstone2018}. Based on the heating and cooling of the atmosphere, and production and losses of neutral hydrogen, the three hydrodynamic regimes \citep[e.g.,][]{ Murray-Clay2009, Bear_Soker2011, Salz2016} have been distinguished - (1) Radiation/Recombination-limited regime, (2) Photon-limited regime and (3) Energy-limited regime. If we concentrate on the production of neutral escaping hydrogen species, then the steady-state continuity equation of neutral hydrogen can be written as following \citet{Lampon2021}:
\begin{equation}
-u\frac{\partial f_{\rm H^0}}{\partial r} + f_{\rm H^+}n_{\rm H^+} \alpha_{\rm H} - f_{\rm H^0}J_{\rm H} = 0,
\end{equation}
where $u$ is the bulk radial velocity, $f_{\rm H^0}$ and $f_{\rm H^+}$ are the mole fraction of neutral hydrogen and ionized hydrogen respectively. $n_{\rm H{^+}}$ is the density of ionised hydrogen; $\alpha_{\rm H}$ and $J_{\rm H}$ recombination and photo-ionisation rates respectively. Based on the production of $f_{\rm H^0}$, we can define the advection rate as $P_{\rm adv} = -u\frac{\partial f_{\rm H^0}}{\partial r}$ and recombination production rate as $P_{\rm rec} = f_{\rm H^+}n_{H^+} \alpha_{\rm H}$. We also need to define another important parameter called {Ionization Front (IF) }- the region of partial ionization where (1- $f_{\rm H^0}) \geq 0.95$. 

Based on the parameters, $P_{\rm rec}$, $P_{\rm adv}$, and IF, the three regimes are classified. In the Radiation/Recombination-limited regime, $P_{\rm rec}/P_{\rm adv} >> 1$ that means advection is negligible and the gas is almost in radiation-recombination equilibrium. The IF region in this case is very narrow compared with the scale of the flow. Heating efficiency-wise, the incident stellar radiation gets significantly reduced by the radiative cooling, and heating efficiency in this case is much less than 0.1. For the Photon-limited regime, the photo-ionization rate is much faster than the recombination rate and the recombination rate is negligible in the upper atmosphere. Hence $P_{\rm rec}/P_{\rm adv} << 1$. In this case, the IF region extends the whole upper atmosphere of the planet. The energy-limited escape is a scenario between the Radiation/Recombination regime and the Photon-limited regime, where advection and recombination rate - none of them are negligible in the upper atmosphere. In this case, the IF is wide but not as wide that it encompasses the entire flow as the photon-limited regime. For the photon-limited and energy-limited regime, the stellar heating is higher as radiative cooling is moderate or negligible, and the heating efficiency is about 0.1.-0.2. All the heating efficiency mentioned here holds true considering the planetary upper atmosphere is hydrogen-dominated. Proper measurements of these hydrodynamic escape regimes are important because different regimes give different mass-loss rates and constraining them would be invaluable for the long-term evolution of the planetary atmosphere. 

\subsection{Non-thermal escape}
Non-thermal escape from exoplanets involves processes in which the velocity of the escaping particles is not related to the temperature of exobase \citep{Tian2015, Gronoff2020}. Non-thermal processes mostly involve the ions and their escape in the presence of electric and magnetic fields. Most of the time, stellar radiation or charge exchange from the solar wind/coronal mass ejections ionizes the neutral constituents of the atmospheres, which further reacts with other neutrals and electrons giving rise to the photochemical escape. Also, the ionized particles can be dragged out of the planet by the solar wind magnetic field. Depending upon the mass, atmospheric composition,  magnetic field of the planet, and orbital distance of the planet from its host star, the non-thermal loss varies. For small rocky exoplanets, photochemical loss becomes important because the exothermic chemical reaction from ionized particles in the atmosphere could provide enough kinetic energy for the escape of the neutral particles in the atmosphere.

The interaction of the upper atmosphere and ionosphere with the solar wind and coronal mass ejections from the host star also significantly contributes to atmospheric mass loss. Neutral atoms in the upper atmosphere get ionized by stellar XUV radiation, charge exchange, and electron impact and are swept away by the magnetic field of solar wind plasma. Pick-up ions, sputtering, and charge exchange contribute significantly to the ion escape to space. Ion escape is one of the major sources of mass loss from the solar-system planets and is also believed to be a great contributor to the mass loss from exoplanets.    

For magnetized planets such as Earth, the pick-up process is less efficient because in that case, the solar and interplanetary electric and magnetic fields remain far from the planet. Ions in the upper atmosphere get trapped in the magnetosphere of the planet and do not necessarily escape the planet. There are two ways the ions could escape in this scenario. For planets with dipolar magnetosphere, the ions can escape to space by the drag of polar outflow. Also, these trapped ions in the planet's magnetic field can eventually escape by charge exchange from the solar/stellar wind plasma. Sputtering is another non-thermal process that is widely believed to be very dominant for escape processes in early Mars and Venus \citep{Luhmann1992, Persson2021} and also very relevant for exoplanet atmosphere \citep{Gu2023}.

\section{Observation of atmospheric escape from exoplanets} \label{sec:observations}
\begin{figure}[htbp!]
\centering
\includegraphics[width=1.0\textwidth]{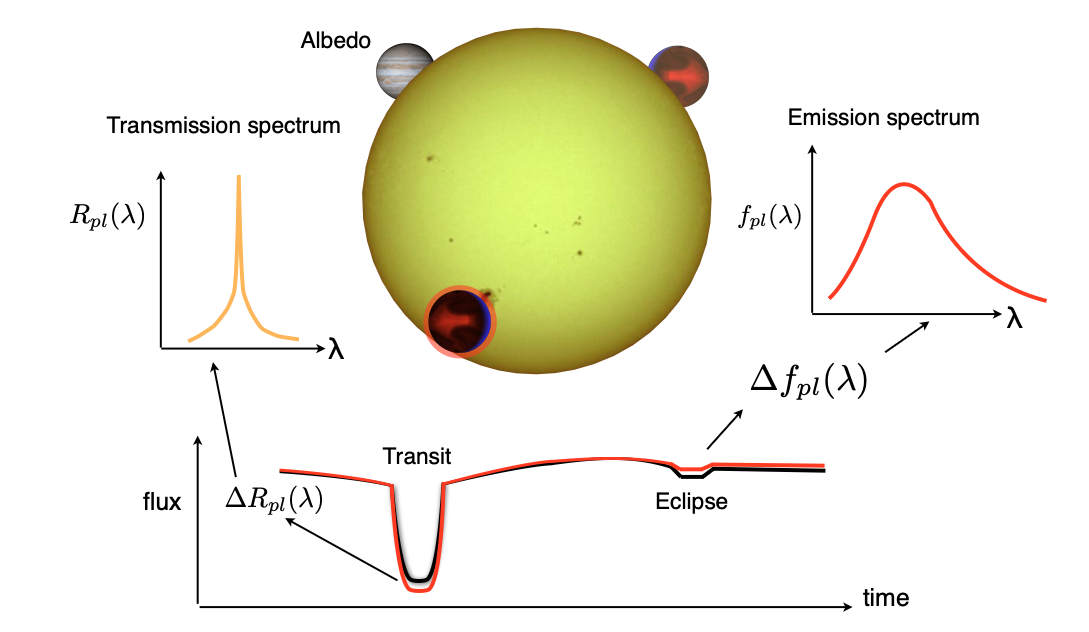}
\caption{A detailed picture of transit spectroscopy when a planet is transiting its host star. Taken from \citet{Sing2018}. The top middle panel shows the scenario of the planet during transit and secondary eclipse. A transmission spectrum as a function of wavelength is shown in the top left. Thermal emission spectrum as a function of wavelength in infrared is measured by the change in the eclipse depth shown in the top right. The transit and eclipse events are shown in the bottom panel. The albedo (fraction of the reflected light from the planet) is measured by observing the secondary eclipse at optical wavelength.}\label{fig:transit}
\end{figure}

Transmission spectroscopy has been used as a very reliable way of measuring atmospheric escape from exoplanets. Exoplanets that are transiting in front of their parent stars provide detailed access to their atmosphere. While exoplanets are in transit, it is possible to separate the planet's signal from that of the parent star allowing us measurement of atmospheric properties of those planets. A transit event (figure~\ref{fig:transit}) allows us to three fundamental measurements for exoplanet characterizations: (1) Transmission spectra - allow us to measure the composition of the atmosphere from the atmospheric absorption features, (2) emission spectra - during the secondary eclipse, the day side average emission of the planet allows us to estimate average temperature of the planet and (3) the phase curve-- by measuring a planet throughout a full orbit, an estimate of the day-to-night temperature contrast can be obtained, which allow us to infer about atmospheric recirculation. As we are interested in atmospheric escape, we focused on the transmission spectra only. 

During a transit event, the planet blocks out some of the stellar light, which we can measure by time-series photometry (see figure~\ref{fig:transit}). By measuring accurately how much light is being blocked by transiting exoplanets, the fundamental properties of those exoplanets such as mass and radius can be determined. The computed fractional flux deficit measured from the stellar light curve (see the black curve in the bottom panel of figure~\ref{fig:transit}), which is proportional to the projected area between the planet and star, gives the planet-to-star radius ratio.  Also, the detailed spectroscopic information of this absorbed light gives us the atmospheric properties of exoplanets. The atmospheric atoms and molecules of planets will absorb and scatter lights at their characteristic wavelengths making the atmosphere opaque on those wavelengths at higher altitudes as shown in the bottom panel of figure-\ref{fig:transit} by the red solid line. Hence measuring the transit spectra in different wavelengths, in principle, we probe different atmospheric compositions in higher altitudes. This is the basic idea of transmission spectroscopy and using this many atmospheric species that are leaving the atmosphere of exoplanets have been detected. 

\begin{figure}[htbp!]
\centering
\includegraphics[width=0.9\textwidth]{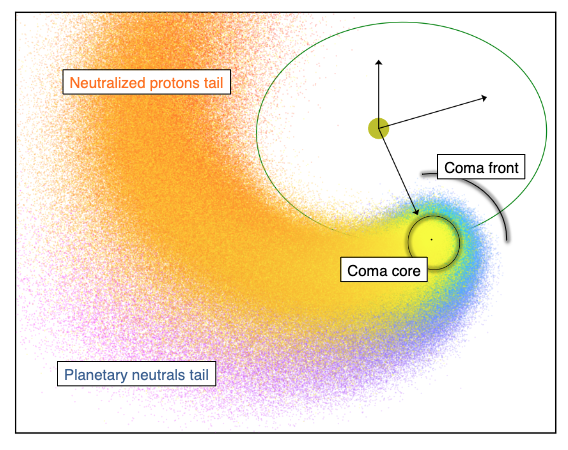}
\caption{A schematic diagram showing the escape of neutral hydrogen from the warm Neptune GJ436b. It shows the planet has an extended tail made of neutral hydrogen and protons. Taken from  \citet{Bourrier2016}}\label{fig:GJ436b}
\end{figure}

As the light elements in the exoplanetary atmosphere such as hydrogen and helium are more susceptible to getting lost in space, the first efforts were rightly put toward the detection of hydrogen in exoplanet atmospheres. The most successful observational probe to establish atmospheric escape comes from the observation of Ly-$\alpha$ transmission from various hot Jupiters and warm Neptunes. \citet{Vidal-Madjar2003} first found evidence of atomic neutral hydrogen absorption in stellar Ly-$\alpha$ by measuring three transits of HD209458b using the  Space Telescope Imaging Spectrograph on board HST. The large absorption $15.0 \pm 4$ \% of stellar Ly-$\alpha$ indicates a large extended envelope of neutral hydrogen, which is beyond the gravitational influence of the planet and hence escaping. This is the first confirmation of neutral hydrogen escaping from the planet HD209458b. Subsequently, many other observations also confirmed the escaping of neutral hydrogen from other exoplanets. For example, \citet{Lecavelier2010, Lecavelier12} found escaping neutral hydrogen in the hot Jupiter HD 189733 b. The warm Neptune GJ436b is a very interesting candidate where an extended tail of escaping neutral hydrogen is observed \citep{Kulow2014, Ehrenreich2015, Lavie2017}. In the figure~\ref{fig:GJ436b}, a depiction of the observed extended tail of escaping neutral hydrogen atmosphere is shown (obtained from \citet{Bourrier2016} ). This planet shows the absorption of stellar Ly-$\alpha$ of $56.3 \pm 3.5$ \%, which shows, so far, maximum absorption of neutral hydrogen representing a huge amount of atmosphere is leaving the planet \citep{Ehrenreich2015}. \citet{Bourrier2018} also reported neutral hydrogen escaping from another warm Neptune GJ3470b with maximum absorption of $35 \pm 7$ \%. There are some limited observations of escaping hydrogen using H-$\alpha$ line also (\citet{Jensen12, Barnes16, Cauley2017}, {\citet{Yan2018, Czesla2022, Bello-Arufe2023}}). 

The stellar Ly-$\alpha$ absorption by planetary atmosphere gives a tremendous opportunity to probe atmospheric escape from the exoplanets, however, note that due to Geo-coronal emission and ISM absorption some part of the Ly-$\alpha$ in the velocity range -50 to +50 km~s$^{-1}$ is not directly observable. As a result, they can not be used to probe the launching region of the planetary outflow as the typical velocity of the gas going through the atmospheric escape is $ \leq 20$ km~s$^{-1}$. The H-$\alpha$ absorption certainly can probe this regime but very limited detection is available due to stellar contamination in the H-$\alpha$ signal. The helium transit line 10830 \AA\ plays an important role in probing this inaccessible region of planetary outflow. 
A major advantage of obtaining helium spectra is that we can observe them from ground-based telescopes. {Recently, there have been significant efforts for the successful detection of He lines from exoplanets from ground based-faciliteis (CARMENES, SPIRou, GIANO, Keck/NIRSPEC) as well as Wide Field Camera 3 on board HST {\citep[e.g.,][]{Allart2018, Oklopcic2018, Spake2018, Zhang2022b, Allart2023, Masson2024, Krishnamurthy24, McCreery2025}}.} Helium detections have been reported in numerous exoplanet targets such as HAT-P-11b \citep{Allart2018}, WASP-107b \citep{Spake2018}, WASP-69b \citep{Nortmann2018}, HD209458b \citep{Alonss-Floriano2019}, HD189733b \citep{Salz2018} and TOI-1807b \citep{Masson2024}. 

{Recently, there have been global surveys \citep[e.g.,][]{Allart2023, Guilluy2023, McCreery2025} of detecting helium from exoplanets using ground-based facilities. To find a definite correlation between mass-loss rate and presence of helium, these homogeneius large-scale analyses is needed. \citet{Allart2023} searched for helium in 11 gas giants with SPIRou but they could not conclude a definite correlation between mass-loss rate and presence of helium, and argued that it still requires larger-scale analysis of the helium triplet. \cite{McCreery2025} analyzed archival helium exoplanet transmission spectra that are detected and publicly available, and concluded that the mass-loss rate follows the energy-limited approximation.} 

In addition to light elements, some other elements including some metals were also found to be escaping from exoplanets \citep[e.g.,][]{Vidal-Madjar04, Fossati2010, Cubillos2020, Ben-Jaffel2022}. \citet{Vidal-Madjar04} from HST measurement detected oxygen(O I) and carbon (C II) in the upper atmosphere of exoplanets that are going through hydrodynamic escape. Recently, ionized iron and ionized magnesium have been observed in hot Jupiters, which must be dragged into the upper atmosphere by the escaping hydrogen and helium \citep[e.g.,][]{Sing2019, Cubillos2020}.

Most of the observational studies of atmospheric escape focused on gaseous exoplanets (e.g., hot Jupiters, warm Neptunes, hot Saturns) because the gaseous exoplanets go through rapid hydrodynamic escape and they are good candidates for atmospheric studies. However, to the best of our knowledge recently there have been no studies that have directly observed atmospheric escape of small rocky exoplanets from observations. Recently \citet{dossantos2019} discussed the possibility of detection of the excess Ly-$\alpha$ absorption signal by hydrogen exosphere around Earth-like planets but we need next-generation spectrographs such as LUVOIR/LUMOS \citep{France2017} to detect it. Also, there is indirect evidence that small rocky exoplanets went through the atmospheric escape in the early evolution times and gave rise to the radius gap \citep{Fulton2017}.

\subsection{Is it possible to segregate different regimes of atmospheric escape from observations?}
As we mentioned in section~\ref{sec2}, the atmospheres of exoplanets go through different types of escape processes, and depending on what kind of process is responsible for their atmosphere to escape, mass-loss rate could be defined. For the solar system planets, there are some studies to understand the transition between Jeans escape and hydrodynamic escape \citep[e.g.,][]{Volkov2011, Erkaev2015}. For the exoplanets, the transition between Jeans escape and hydrodynamic escape {has not been} studied at all from observations. However, \citet{Lampon2021} successfully distinguished the three hydrodynamic escape regimes that theoretical models predicted (see section~\ref{hydroescape}) by observing transmission helium triplet lines and Ly-$\alpha$ in gaseous exoplanets. 

The high-velocity neutrals ($\simeq$ 200 km~s$^{-1}$) observed in Ly-$\alpha$ during the transit event from gaseous exoplanets \citep[e.g.,][]{Lecavelier12, Ehrenreich2015} are signature of escape which are most likely driven by stellar wind/coronal mass ejection interaction \citep{Carolan2020, Hazra2022}. Since the XUV-driven thermal escape can not accelerate neutral particles {typically} beyond 20-30 km~s$^{-1}$ {\citep{Murray-Clay2009, Salz2016, Kubyshkina2018b, Allan2019,Hazra2020, Schreyer2024, Rosener2025}}, high-velocity neutrals are due to non-thermal escape from the planet and can be differentiated from low-velocity thermal escape ($\leq$ (20 - 30) km~s$^{-1}$).

\section{Numerical models to understand atmospheric escape}\label{sec:numerical_model}
Exoplanet research is a relatively new field only 2 decades old. After the first detection of atmospheric escape from the hot Jupiter HD209458b in 2003 \citep{Vidal-Madjar2003}, different groups started to develop theoretical models to understand the atmospheric escape from exoplanets. The early efforts were based on 1D models for both gas giants and rocky terrestrials exoplanets \citep[e.g.,][]{Yelle2004, Tian05, Garcia-Munoz2007, Murray-Clay2009, Guo2011, Koskinen13, Johnstone2015, Guo2016, Kubyshkina2018, Johnstone2018, Allan2019, Hazra2020}. With time, 1D escape models have been enriched with detailed complicated physics, and still, new things are being added to those models. However, some of the aspects can not be modeled using simple spherically symmetric models. For example, the observed extended tail-like structure \citep{Ehrenreich2015} of transiting exoplanets can not be modeled using 1D models. The day-night asymmetry, changes in temperature between day and night sides, and the complex magnetospheric structure could only be modeled using 3D models. The 3D models of atmospheric escape have also been started a decade ago and they are now becoming very sophisticated in explaining various observational results of atmospheric escape from gas giants \citep[e.g.,][]{Tripathi2015, Villarreal2014, Khodachenko2015, Christie2016, Carroll-Nellenback2017, Villarreal2018, Esquivel2009, McCann19, Carolan2020, Hazra2022, MacLeod2022}, as well as from terrestrial exoplanets \citep[e.g,][]{Dong2017, Garcia-sage2017, Dong2018, Dong2019}. In the next few sections, I will discuss 1D models and 3D models for atmospheric escape including their results.     

\subsection{1D escape models}
The first kind of hydrodynamic escape model was developed by \citet{Yelle2004} for the giant planets that are orbiting the host star in very close proximity within 0.01 to 0.1 AU. They solve the usual fluid equations for hydrodynamic escape (see section~\ref{hydroescape}) with stellar heating, adiabatic cooling, and photochemical reactions. Their model included neutral species H$_2$, He, and H and ionized species H$^+$, H$^+_2$, H$^{+}_3$, He$^+$, and HeH$^+$ and solved the continuity equation and diffusion equation along with momentum, energy equations for the composition of the upper atmosphere of giant planets. The mass-loss rate (escape rate) 10$^8$ g~s$^{-1}$ calculated from their model was 100 times smaller than the mass-loss rate estimation of \citet{Vidal-Madjar2003} from Ly-$\alpha$ observation. I shall use mass-loss rate or escape rate interchangeably in this whole review if not otherwise mentioned.  

The discrepancy in the mass-loss rate for the giant planets obtained from the model of \citet{Yelle2004} and observations instigated further study on this topic. \citet{Tian05} used updated two-dimensional heat deposition studying hydrodynamic escape of hydrogen-rich atmosphere. They found the atmospheric mass-loss rate of $6 \times 10^{10}$ g~s$^{-1}$, which is in good agreement with the observed value from HD209458b despite the assumption of solar equivalent EUV radiation from the host star. With a sophisticated molecular diffusion model, \citet{Garcia-Munoz2007} was also able to find a mass loss rate comparable to the observation. \citet{Murray-Clay2009} have added self-consistent radiative transfer for the stellar EUV heating and radiative cooling and computed the hydrodynamic equations for escape calculation. In addition to the calculation of a mass-loss rate of neutral hydrogen, they calculated the synthetic transit spectra for direct comparison with the observed transit calculation of Ly-$\alpha$ for HD209458b \citep{Vidal-Madjar2003}. 

\begin{figure}[htbp!]
\centering
\includegraphics[width=0.475\textwidth]{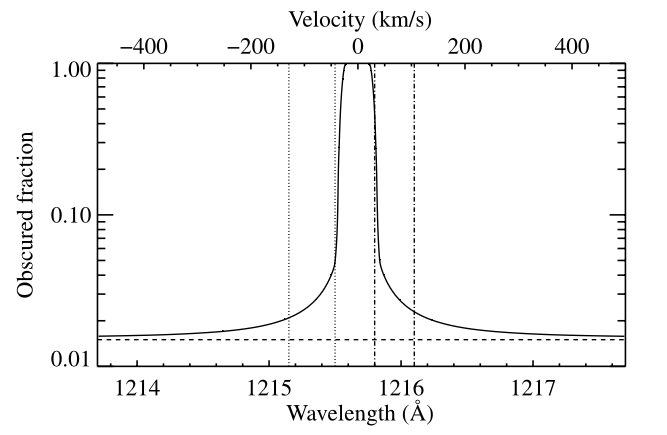}
\includegraphics[width=0.45\textwidth]{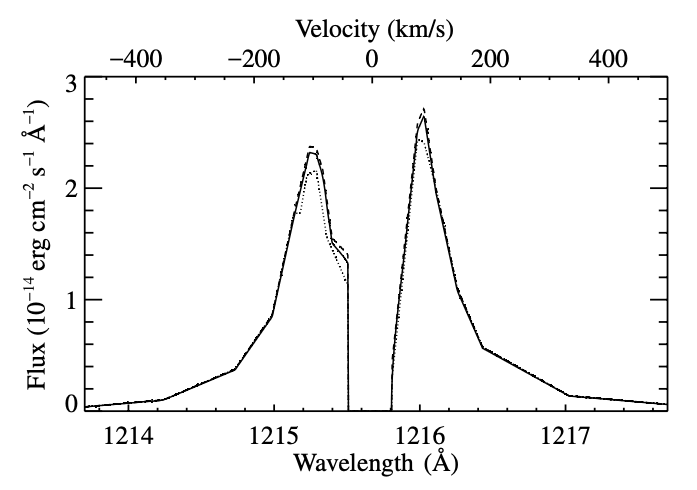}
\caption{left: Synthetic Ly-$\alpha$ transit spectra computed from the escaping hydrogen atmosphere of HD209458b. All required density, velocity, and temperature for synthetic spectra are computed using the escape model of \citet{Murray-Clay2009}. Right: Direct comparison of simulated spectra with the observed spectra after correcting for the interstellar absorption. Taken from \citet{Murray-Clay2009}}\label{fig:transit_murray-clay}
\end{figure}

In the left panel of figure~\ref{fig:transit_murray-clay}, I show the theoretically computed synthetic spectra for neutral hydrogen absorption in the Ly-$\alpha$ line from \citet{Murray-Clay2009}. The plot shows the amount of stellar Ly-$\alpha$ obscured by the planetary atmosphere ($e^{-\tau}$), which is estimated by calculating line-of-sight optical depth integrated up to a maximum altitude of 10R$_p$, where R$_p$ is the radius of the planet. The horizontal dashed line shows the geometrical visible continuum. Two sets of dotted and dot-dashed lines show the Doppler velocity of escaping material near $\pm 100$ km~s$^{-1}$. In the right panel, the dashed line and dotted line show the observed out-of-transit and in-transit spectra. The line-core (-42 to +32 km~s$^{-1}$) is set to zero where interstellar absorption is strong. Using the theoretical absorption (as shown in the left panel of this figure), the in-transit spectrum is overplotted using a solid line. It has been found that the theoretical model of \citet{Murray-Clay2009} was not able to explain the observed 15\% absorption in the high-velocity neutrals. The model spectra showed a decrement of 3\% of stellar flux in the observed in-transit line spectrum. A plausible reason for this less absorption in the high-velocity neutrals is the absence of stellar wind interaction with the planetary outflow, which needs multidimensional modeling (see next section~\ref{sec:3Dmodel}). Based on the relations between both stellar heating and radiation cooling with mass-loss rate, \citet{Murray-Clay2009} found that for the intense XUV flux from the star, the atmospheric escape is radiation/recombination limited, and for low XUV flux, it is energy-limited.  

The above models explained so far have assumed the planetary atmosphere is dominated by hydrogen-helium. However,\citet{Vidal-Madjar04} found other species in the atmosphere are escaping too. \citet{Koskinen13} added heavier atoms and ions in their calculation of hydrodynamic escape and discussed the possibility of the presence of heavier neutral and ionized atoms in the upper atmospheres. 

\begin{figure}[htbp!]
\centering
\includegraphics[width=0.99\textwidth]{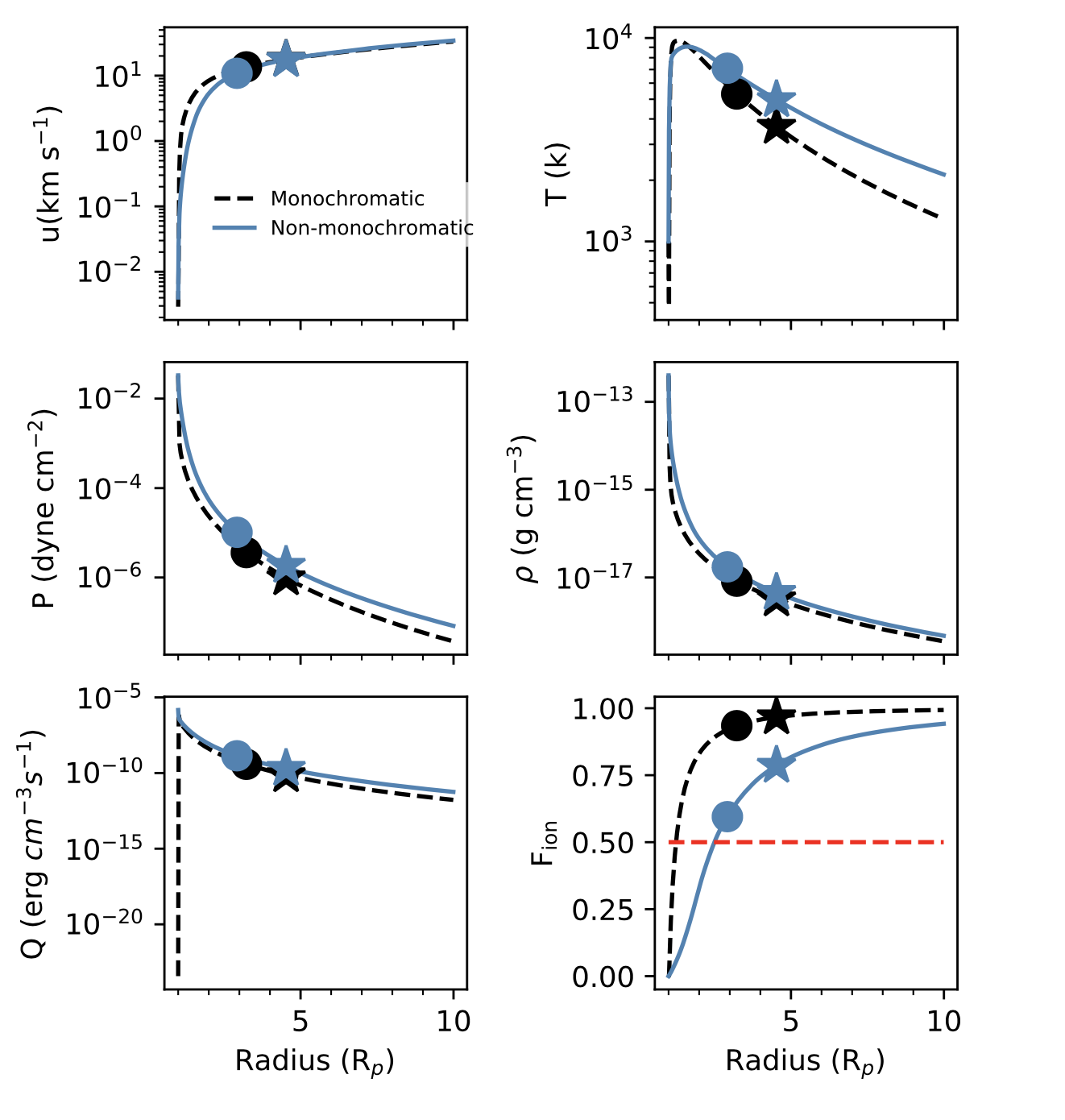}
\caption{Behaviour of upper atmosphere of a hot Jupiter from the simulation of hydrodynamic escape for monochromatic and non-monochromatic radiative transfer of stellar radiation. The black dashed line and solid blue lines show monochromatic and non-monochromatic cases respectively. The velocity (u) and temperature (T) of the escaping materials are shown on the left and right of the top panel. Similarly, pressure (P) and mass density ($\rho$) are presented in the middle row. The total volumetric heating rate and ionization fraction of hydrogen are given in the left and right of the bottom row, respectively. This figure is adopted from \citet{Hazra2020}. The filled circles show the sonic point and star markers show the Roche lobe boundary for the star-planet system that they considered.}\label{fig:escape_hazra20}
\end{figure}

One of the major approximations in the \citet{Murray-Clay2009} while implementing radiative transfer is the monochromatic approximation - all stellar spectral energy distribution (SED) is concentrated on a single wavelength corresponding to energy of 20 eV. This approximation was reasonable given the difficulty in observing the full SED from the host stars. However, with new techniques and machine learning, it is possible to obtain SED from stars \citep{Sanz-Forcada11, Linsky2014, France2016, France2020, Behr2023}. \citet{Hazra2020} considered a full SED from the host star and studied the effect of full SED on the hydrogen-dominated atmosphere. They assumed a hypothetical hot Jupiter orbiting our present Sun and obtained full solar SED in the XUV regime, which was driving the atmospheric escape. The comparison of physical properties of the atmosphere for both monochromatic approximation and full SED implementation are shown in figure~\ref{fig:escape_hazra20}. The results for both cases are not very different except for the temperature profile and the ionization fraction. In the right panel of the bottom row, half of the ionization fraction is marked using a red horizontal dashed line. Interestingly, the ionization is stronger for the monochromatic case which will in principle predict less Ly-$\alpha$ absorption due to less neutral hydrogen but can not be probed by observation as the line core of Ly-$\alpha$ is not visible.   

{With the recent discovery of helium as an excellent probe for the escaping atmosphere, there are some efforts \citep[e.g.,][]{DosSantos2022,Biassoni2024,Linssen2024, Rosener2025} in developing 1D hydrodynamic escape model that interpret helium observations and retrieve the atmospheric mass-loss and temperature of upper atmosphere. \citet{DosSantos2022} developed an open source python code {\it p-winds} that couples a purely upper atmosphere H + He Parker wind model with ionization balance, ray tracing and radiative transfer. By fitting {\it p-winds} models with observed helium spectra, they are able to estimate mass-loss from the planet but the model results depend on the hydrogen mass fraction. In a similar line, \citet{Linssen2024} developed a model called {\it sunbather} which couples {\it Cloudy} interface for doing photoionsiation and radiative transfer calculation with {\it p-winds}. The open source code {\it Pluto} has been also used to simulate upper atmosphere by coupling with {\it Cloudy} to develop an open source code {\it pyTPCI} for simulating outflows and their signature in helium and metal lines \citep{Rosener2025}.}

For the small terrestrial exoplanets, the computation of atmospheric escape involves much-complicated photochemistry, heating, and radiative cooling. \citet{Johnstone2018} considered heating from the absorption of stellar X-ray, UV, and IR radiation. They also included heating from exothermic chemical reactions, electron heating from collisions with non-thermal photoelectrons, and Joule heating. For the cooling, cooling from infrared emission by several species is included. The thermal conduction and energy exchanges between the neutral, ion, and electron gases are included in their model. Altogether, they considered a chemical network that includes $\sim$ 500 chemical reactions including 56 photoreactions, eddy and molecular diffusion, and advection. This sophisticated model has been validated by studying the thermospheric structure of Earth and Venus. For modeling the upper atmosphere of terrestrial planets and understanding atmospheric escape from them, this type of model will play a significant role. 

Most of the 1D models that we discussed above dealt with the hydrodynamic escape from the gas giants and terrestrial exoplanets, as hydrodynamic escape is quite significant for evolution. In addition, most of the observations available from the transit calculation, are for those planets which are very close to the host star. Being in close proximity to the host star, it is expected that planets would go through extreme hydrodynamic mass loss. For solar system planets, the Jeans parameter (see equation~\ref{eq:jeans} in section~\ref{sec:jeans}) is a very good indicator to understand if the planet is going through the Jeans escape or hydrodynamic escape \citep[e.g.,][]{Volkov2011}. Generally, if a Jeans parameter is larger than $\sim$ 2-3.6, a transition from hydrodynamic escape to Jeans escape occurs. There were not many studies on exoplanets to understand which type of escapes are going on exoplanets until recently \citet{Guo2024} analyzed a large sample of approximately 2700 planets orbiting G and M-type stars, which are receiving XUV flux and under the influence of their host stars. By defining a new Jeans parameter ($\lambda^{*}_{\rm esc}$) which takes into account the effect of tidal force (tidal force is not very important for solar system planets as they are far from the Sun), a new limit on it is set for the transition of hydrodynamic escape to Jeans escape. They reported that, unlike solar system planets, Jeans escape does not occur for close-in exoplanets if the $\lambda^{*}_{\rm esc}$ is larger than 3. Even XUV-driven hydrodynamic escape can happen if $\lambda^{*}_{\rm esc}$ is more than 100.   

\subsection{Multi-dimensional escape models} \label{sec:3Dmodel}
The 1D escape models were successful in reproducing many features of atmospheric escape qualitatively including mass loss rates but the details of atmospheric escape such as day-night asymmetry, comet-like structure, and presence of high-velocity neutrals ($\geq$ 100 km~s$^{-1}$) in the upper atmosphere were not possible to capture in 1D. The first kind of multidimensional atmospheric escape simulation was started by \citet{Stone2009} in two dimensions. This model used a simplified treatment of radiative transfer and thermodynamics. The multi-dimensionality of the model produces significantly different results than spherically symmetric models with the same parameters. The mass-loss rate was reduced 4 times in these models than in spherical wind and the sonic point is close to the planetary surface.

The 3D simulations of the atmospheric escape process with a sophisticated treatment of radiative transfer and detailed microphysics have been performed by many groups till now \citep[e.g.,][]{Tripathi2015, Villarreal2014, Khodachenko2015, Christie2016, Carroll-Nellenback2017, Dong2018, McCann19, Carolan2020, Hazra2022}.  Here we mention a few. \citet{Tripathi2015, McCann19} performed atmospheric escape simulations using open-source Athena code and  \citet{Carroll-Nellenback2017, Debrecht2020} used Astrobear code. \citet{Hazra2022} developed a 3D radiation-hydrodynamic escape model based on open source BATS-R-US code. They simulated the upper atmosphere of hot Jupiter HD189733b based on self-consistent radiative transfer of stellar heating and radiative cooling. They solve the hydrodynamic equations in 3D cartesian grids that include adaptive mesh refinement as given below:   

\begin{equation}\label{eq:cont_3D}
\frac{\partial{\rho}}{\partial{t}} +\nabla \cdot \rho {\vec{ u}} = 0,
\end{equation}
\begin{equation}\label{eq:mom_3D}
\frac{\partial(\rho\vec{u})}{\partial t} + \nabla \cdot [\rho \vec{u} \vec{u} + P_TI] = \rho\bigg( \vec{g} -\frac{GM_{*}}{(r-a)^2} \hat{R} - \vec{\Omega} \times (\vec{\Omega}\times\vec{R})-2(\vec{\Omega} \times \vec{u}) \bigg),
\end{equation}
%
\begin{eqnarray}\label{eq:energy_3D}
\frac{\partial }{\partial t} \left( \frac{\rho u^2}{2} + \frac{P_T}{\gamma -1} \right) + \nabla \cdot [\vec{u}(\frac{\rho u^2}{2} + \frac{\gamma P_T}{\gamma -1} )] = \rho \bigg( \vec{g} -\frac{GM_{*}}{(r-a)^2} \hat{R} - \vec{\Omega} \times (\vec{\Omega}\times\vec{R}) \bigg) \cdot \vec{u} ~\nonumber \\ 
+{\cal H}-{\cal C}~~.~~~~~~
\end{eqnarray}

\begin{figure}[htbp!]
\centering
\includegraphics[width=0.99\textwidth]{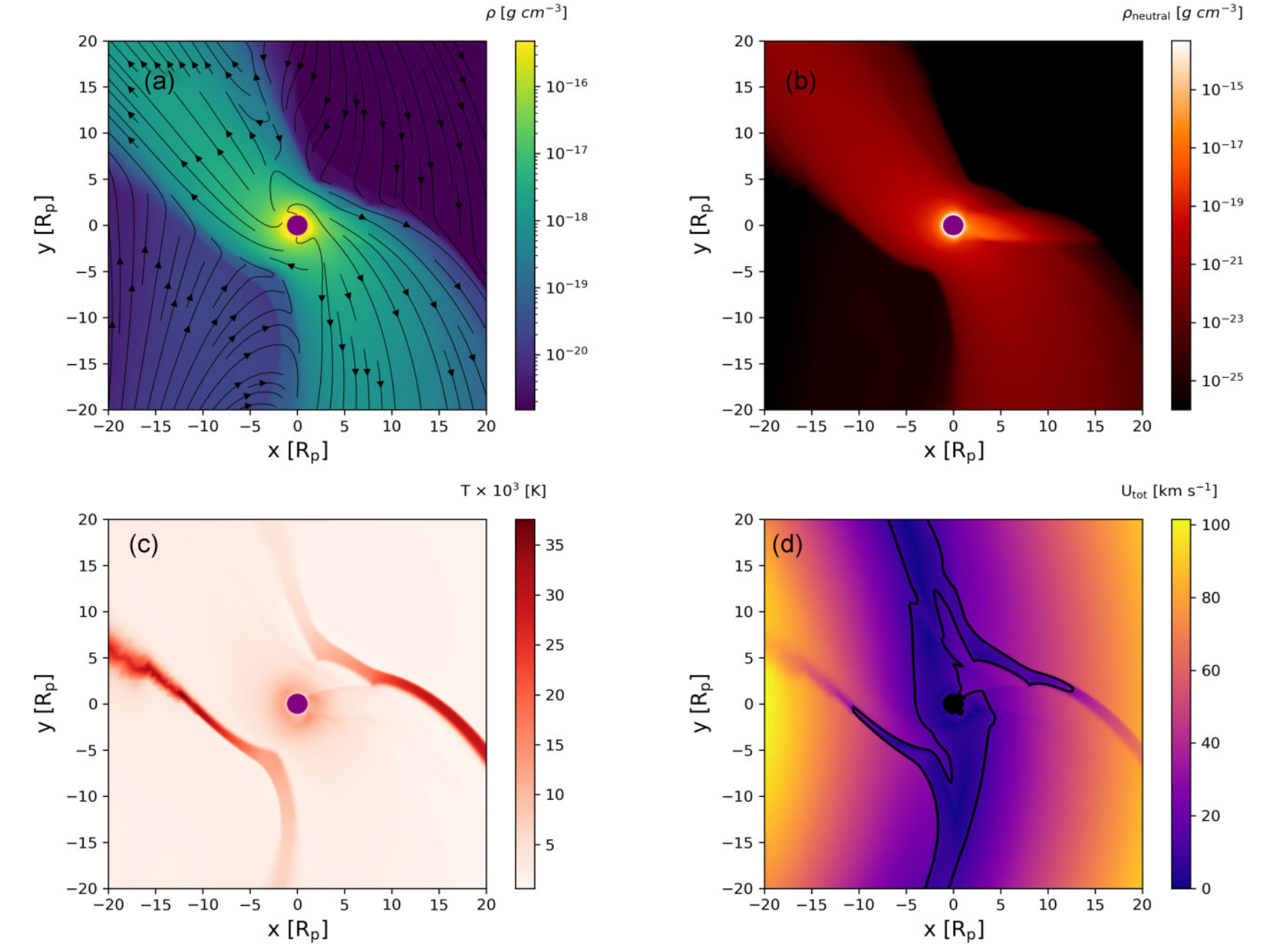}
\caption{(a)Total density (neutral + ions + electrons) of outflowing materials around the planet. Velocity streamlines are shown in black contours. (b) The neutral hydrogen profile, (c)The temperature distribution, and (d) The total velocity of the outflowing materials including the sonic surface. The planet is in the center of the grid shown using the purple circle. All plots are in the orbital plane. In these all simulations, the host star is kept outside of the simulation domain at the negative axis. Taken from \citet{Hazra2022}}\label{fig:atmosphere_3D}
\end{figure}

All notation is usual and details are given in \citet{Hazra2022} and we are not repeating here. {However, it is worth mentioning some comments on the numerical resolution. Any numerical simulation should be able to resolve the scale height of the atmosphere to capture the underlying atmospheric variations. For example, the hydrostatic pressure scale height $H$= $kT/(m_H g)$ for a typical hot jupiter ($\sim 0.7M_J$) = 0.1R$_p$ \citep{Murray-Clay2009}. Hence the numerical resolution in the simulations should be higher than 0.1R$_p$. Usually, the computational volume around the planet is taken much larger than the planetary environment to minimize the boundary effect. For HD189733b, the Cartesian simulation box considered by \citet{Hazra2022} has a rectangular grid with x = [-20, +40]R$_p$ , y = [-40, +40]R$_p$ , and z = [-32,+32]R$_p$. However, for HAT-P-11b, as the magnetotail is expected to be longer from the condition of stellar wind, a larger computational domain has been used \citep[for details see][]{Ben-Jaffel2022}. The inner boundary for atmospheric escape simulation are fixed at the distance where the UV photons are absorbed (optical depth $>> 1$). This is close to the surface of the planet (approximately at a height of 1.1R$_p$) with atmospheric pressure $\sim 1$ nanobar at the base of the wind. The visible radiation from the star is absorbed at pressure closer to 1 bar at the surface and it takes about 0.1R$_p$ for atmospheric pressure to be reduced from 1 bar to 1 nanobar. In this lower atmosphere (higher pressure zone), we need to generally employ Global Circulation Models to understand the behaviour of the atmosphere.} The stellar heating using radiative transfer is implemented in the heating term ${\cal H}$ in the equation~\ref{eq:energy_3D}. In this simulation, Coriolis force and centrifugal forces are also included. The atmospheric properties of the planet due to stellar heating are shown in the figure~\ref{fig:atmosphere_3D}. The total density, neutral hydrogen density, temperature, and total velocity of outflowing materials are shown in the four panels (a), (b), (c) and (d) of figure~\ref{fig:atmosphere_3D} respectively. The 3D model shows morphologically different results than 1D models as expected. The atmosphere of the planet moves in a clockwise direction due to the Coriolis force. The velocity streamlines by black contours (panel (a) of figure~\ref{fig:atmosphere_3D}) show this clockwise movement of the outflowing material clearly. One interesting thing to notice in this atmospheric profile is that total density does not show significant day-night asymmetry as is expected from the stellar heating. However, the neutral density shown in panel (b) has a distinct shadow on the night side of the planet due to no ionization of planetary material.  The interplay between tidal force and Coriolis force on the planetary material in different directions creates shocks which are clear from the high-temperature regions in panel (c). In panel (d), the total velocity with the sonic surface is shown. Because of the competition among tidal force, Coriolis force, XUV heating, and gravitational force, the sonic surface varies from the day side to the pole to the night side. This shows the importance of multidimension in simulating the planetary outflow as planetary outflow originates at different distances around the planet.   

The estimated mass-loss rates from the 3D atmospheric escape models \citep{Tripathi2015, Carroll-Nellenback2017, McCann19} are comparable with observed values \citep{Vidal-Madjar2003, Lecavelier2010, Lecavelier12, Bourrier2016}. For the planet HD189733b, the estimated mass-loss rate from 3D model of \citet{Hazra2022} is $6 \times 10^{10}$ g~s$^{-1}$. This value is comparable with several existing models \citep{Guo2011, Guo2016, Salz2016} and observations \citep{Lecavelier12}.

\subsubsection{Interaction of stellar wind with planetary upper atmosphere and its importance on atmospheric escape}\label{sec:sw}
\begin{figure}[htbp!]
\centering
\includegraphics[width=0.99\textwidth]{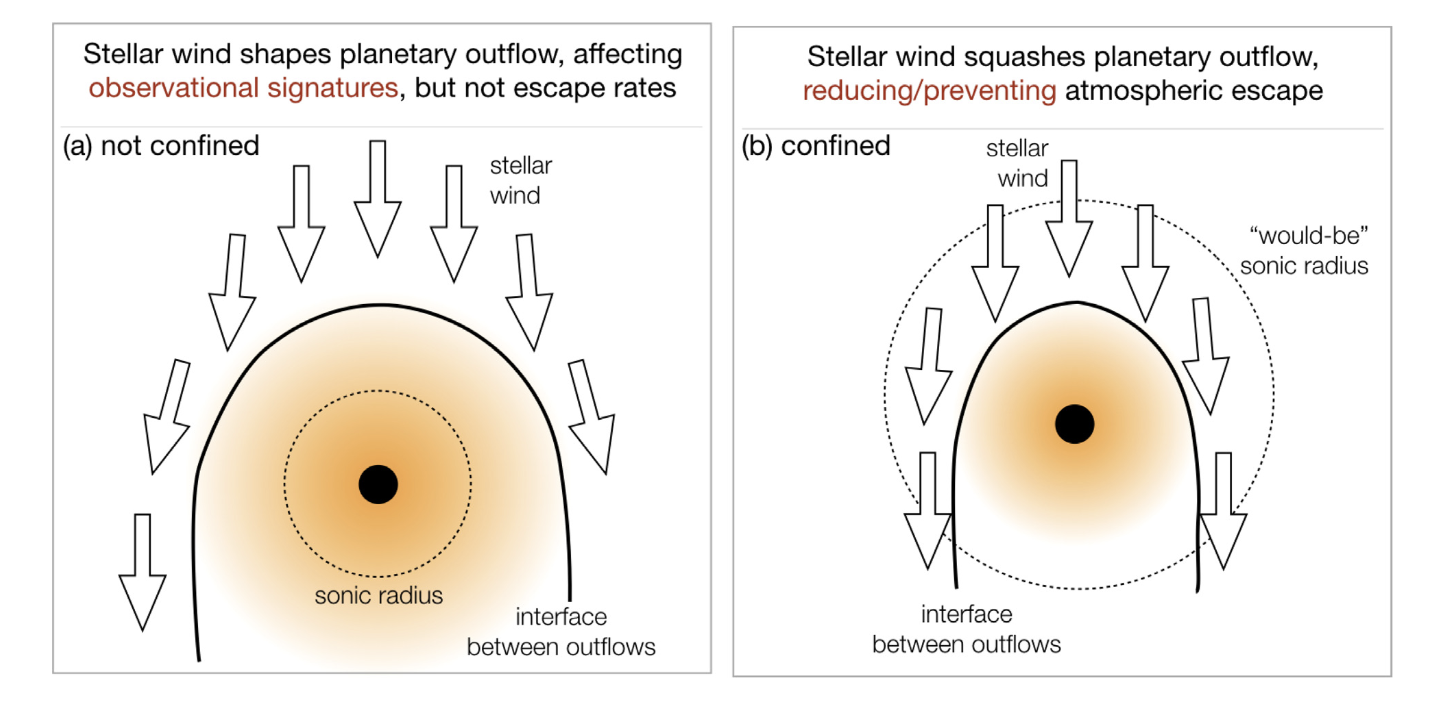}
\caption{Schematic diagram of confinement of planetary outflow by stellar wind. Left panel: Not confined - the case where stellar wind ram pressure is weaker than the ram pressure of planetary outflow. Right panel: Confined- the case where stellar wind ram pressure is stronger than the ram pressure of planetary outflow. Figure is taken from \citet{Vidotto2020}}\label{fig:stellar_wind_confinement}
\end{figure}

The hydrodynamic outflow that originates from the planet due to incident stellar XUV radiation interacts further with the plasma environment of the host star. The interaction between stellar wind and planetary outflow due to atmospheric escape shapes up the upper atmosphere of the planets. \citet{Vidotto2020} studied the effect of stellar wind on the atmosphere of close-in exoplanets in detail. The competition between stellar wind ram pressure at the orbital distance of the planet and the ram pressure of the exoplanetary outflow at the sonic point decides whether the atmospheric mass loss from the planet will be affected. In figure~\ref{fig:stellar_wind_confinement}, the scenario is explained. If the stellar wind ram pressure at the orbital distance is stronger than the ram pressure of the planetary outflow at the sonic point, then the planetary outflow gets confined from the supersonic outflow to the subsonic breeze changing the escape rate. This situation is shown in the right panel of figure~\ref{fig:stellar_wind_confinement}. If the situation is another way i.e, the ram pressure of planetary outflow at sonic pint is stronger than the stellar wind ram pressure, planetary outflow remains unconfined and escape rate does not get affected (see the left panel of figure~\ref{fig:stellar_wind_confinement}). This is a general consensus about the interaction of stellar wind and its impact on the planetary atmosphere. However, this is based on 1D calculation and purely hydrodynamic interaction between stellar wind and planetary outflow.   

\begin{figure}[htbp!]
\centering
\includegraphics[width=0.99\textwidth]{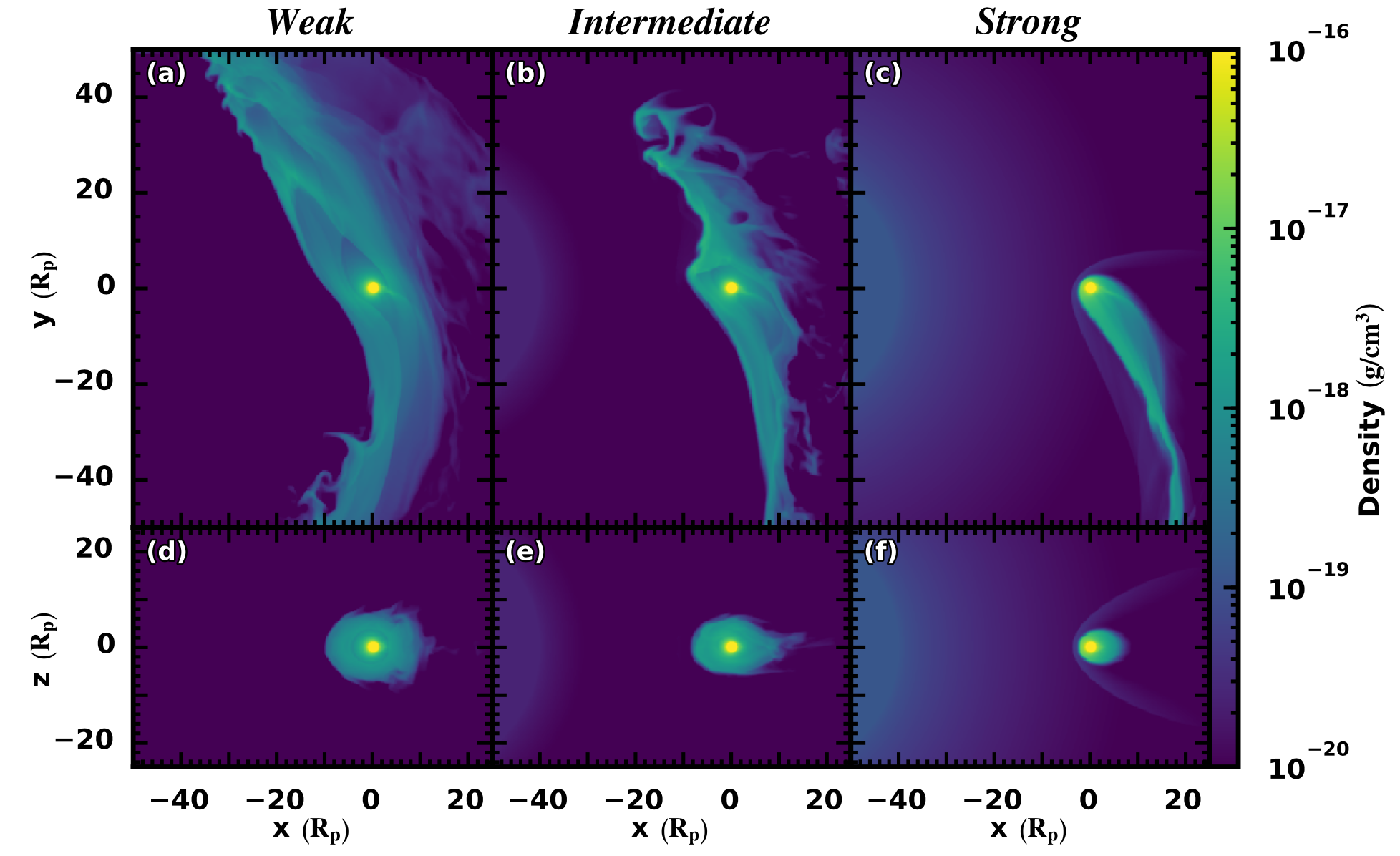}
\caption{Total density of planetary outflow interacting with the weak (first column), intermediate (second column), and strong (third column) stellar wind . The upper row shows a density pattern in the orbital plane (z=0) and the lower panel shows the same in the polar plane (y =0). The figure is taken from \citet{McCann19}}\label{fig:McCan_stellarwind}
\end{figure}

{Some 3D escape models explore the stellar wind interaction with the planetary outflow \citep[e.g.,][]{Carroll-Nellenback2017, McCann19, Carolan2020, Carolan2021a, Wang2021a, MacLeod2022, Hazra2022, Mitani2022}}. In a 3D simulation, \citet{McCann19} found that stellar wind indeed can confine the atmospheric escape from the planetary atmosphere if the stellar wind is sufficiently strong. In figure~\ref{fig:McCan_stellarwind}, we show a set of simulations from their paper, which depicts the interaction of stellar wind of three different strengths with planetary outflow that has a constant escape rate for three stellar wind cases. The upper panel shows the orbital plane (z = 0) and the lower panel shows vertical slices perpendicular to the star-planet axis (y = 0). As we expected, the weaker stellar wind in column 1 ((a) and (d)) could not confine the planetary outflow and did not affect the mass loss rate. The intermediate stellar wind could partially confine the planetary outflow (see second column, (b) and (e)). Finally, the strong stellar wind can confine the planetary outflow reducing it to a breeze (column 3, (c) and (f)) of figure~\ref{fig:McCan_stellarwind}. The stronger stellar wind ram pressure can suppress the whole day-side outflow and shape the atmosphere of the planet into a comet-like structure, which has been observed already \citep{Ehrenreich2015, Bourrier2016}. A similar conclusion was also drawn by \citet{Carolan2020, Carolan2021a}. These results are generally applicable to the hot and warm gaseous planets.     

\begin{figure}[htbp!]
\centering
\includegraphics[width=0.99\textwidth]{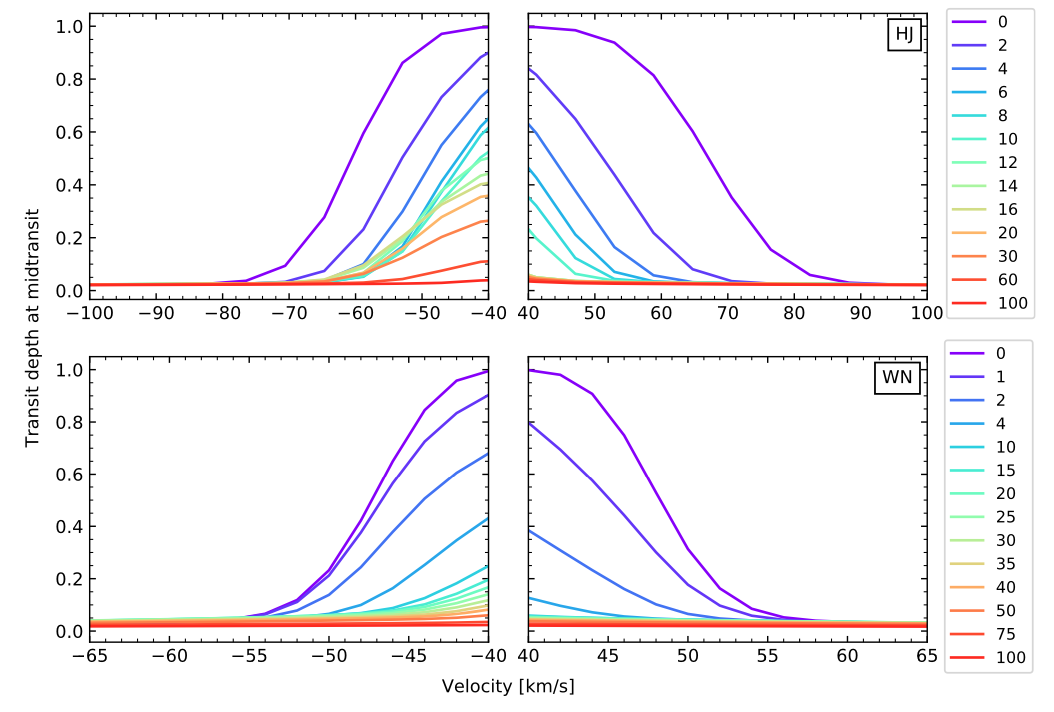}
\caption{Ly-$\alpha$ transit spectra from atmospheres of a hot Jupiter (top panel) and a warm Neptune (bottom panel) for different strengths of stellar wind from their host stars. The left and right panels show the blue wing [-40 to -300 km~s$^{-1}$] and red-wing [+40 to +300 km s$^{-1}$] parts of the transit spectrum. The different colors represent the variation of stellar mass-loss rate from 0 to 100 $\dot{M}_\odot$. The gap in figure from -40 to + 40 km~s$^{-1}$ is kept as the line center gets contaminated by interstellar absorption and geocoronal emission. This figure is taken from \citet{Carolan2021a}. }\label{fig:transit_carolan}
\end{figure}

The interaction of stellar wind with the planetary atmosphere not only changes the mass-loss rate but also changes the transit signature. {Most of the simulations (mentioned above) of atmospheric interaction with the stellar wind have been performed so far assuming a hydrogen-rich atmosphere, except a few recent study \citep{Wang2021a, Wang2021b, Khodachenko2021,MacLeod2022} that modeled the hydrogen-helium atmosphere}. The Ly-$\alpha$ transit signatures have been synthetically generated from these 3D models for different strengths of stellar wind. For details of transit signature calculation from the 3D simulation, please see section~4 of \citet{Carolan2021a}. \citet{Carolan2021a} computed the Ly-$\alpha$ transit signatures for different stellar wind strengths for a hot Jupiter and a warm Neptune, which is shown in figure~\ref{fig:transit_carolan}. The stellar wind strength is determined by its mass loss rate. The higher the mass loss, the stronger the wind is. They have performed a series of simulations with varying mass-loss rates of stellar wind from 0 to 100$\dot{M}_\odot$. {The figure clearly shows that as the stellar wind strength increases, it confines the atmospheric escape but increases the asymmetry in transit absorption spectra. As stellar wind strength increases, transit depths in high-velocity blue wing (integrated transit depth over the velocity range of [-300, -40 km~s$^{-1}$]) and red wing [+40, +300 km~s$^{-1}$) decrease for both hot Jupiter and warm Neptune. However, the transit depth in red wing decreases more than the blue wing as with increasing stellar wind strength the day-side outflow gets confined completely. The different colors show the different mass-loss rates of stellar wind.}  


{It has been found that the weak stellar wind can not change the mass-loss rate much as the way the stronger stellar wind does, but it changes the transit signatures. \citet{Carolan2021a} reported that for stellar wind with mass loss rates $<$ 10$\dot{M}_\odot$, the atmospheric escape does not change much but blue wing absorption in Ly-$\alpha$ changes approximately 10\% for hot Jupiter to 3\% in warm Neptune. \citet{MacLeod2022} computed the synthetic transit spectra for the metastable helium from their simulation of stellar wind interaction with the planetary atmosphere. They simulated three cases A, B and C with weak, intermediate, and strong stellar wind similar to the cases shown in figure~\ref{fig:McCan_stellarwind}. The simulated light curve of excess metastable absorption shows that for the case of weak and intermediate stellar wind (cases A and B in their paper), the helium absorption is not significantly different than the case with no stellar wind. For the strong stellar wind, they found a large absorption in the helium triplet line (see figure-4 of \citet{MacLeod2022}). This indicates that the stellar wind confines the atmospheric escape with concentrated planetary outflowing materials close to the planet, which increases the optical depth and hence the helium transit depth.} 


The above argument of the interaction of stellar wind with the planetary atmosphere would be appropriate for most of the hot and warm gaseous planets that are going through hydrodynamic escape. For the small low-mass terrestrial planets with hydrogen-helium envelope, in the early phase of their evolution, the strong hydrodynamic escape in principle would be confined or not, will be decided by the stellar wind strengths. Atmospheric escape of mid-aged rocky exoplanets like our solar system planets, which do not go through strong hydrodynamic escape or lost their primary atmosphere, will be largely affected by stellar wind by other processes (e.g., pick-up ion process, ion-loss, sputtering). \citet{Dong2017} and \citet{Dong2018} simulated the interaction of stellar wind with the atmosphere of Proxima-Centauri b and Trappist-1 planets, respectively. Using their numerical model, they found that stellar wind increases the atmospheric loss from those planets significantly. The ion-loss (O$^{+}$, O$_2^{+}$, CO$_2^{+}$) for all the seven Trappist-1 system planets was higher when a strong dynamic pressure of the stellar wind is considered in comparison to a lower dynamic pressure (see Table-1 of \citet{Dong2018} for details). This shows a very interesting conflict of atmospheric mass loss from the gaseous planets and terrestrial exoplanets. The gaseous exoplanets or rocky exoplanets with hydrogen-helium envelops show a decrease in mass-loss rate due to strong stellar wind, whereas rocky exoplanets with present-day Mars or Venus compositions show an increase of mass-loss rate due to high stellar wind ram pressure. Of course, different physical processes must be acting on different parameter regimes but understanding this dichotomy properly would be very suitable for the atmospheric escape process from exoplanets as well as habitability.    

\subsubsection{Effect of stellar and planetary magnetic fields on escape}
In the solar system planets, the magnetic field plays an important role in the atmospheric escape processes \citep{Ma2004, Li2021}. This is very likely that the magnetic field would play a role in the atmosphere escape process in the exoplanets. Efforts to understand the role of the magnetic field in atmospheric escape started by a few groups in both 2D \citep[e.g.,][]{Owen2014, Trammell2014} and 3D \citep[e.g.,][]{Cohen2011, Matsakos2015, Cohen2015, Carolan2021}. In a series of two-dimensional simulations of hot Jupiter's upper atmospheres with included planetary magnetic field, \citet{Trammell2014} found that planetary outflow has a three-zone structures - polar dead zone, mid-latitude wind zone, and equatorial dead zone. The magnetic pressure associated with the imposed dipolar magnetic field does not allow the material in the equatorial region to escape resulting in a dead zone. In the mid-latitudes, material can escape by following open field lines. Interestingly, they found polar dead zones too which are created by stellar tides. The central result was that with the increasing strength of the planetary magnetic field, the escape rate of the planetary atmosphere decreased. {The reason is straightforward. With the increasing magnetic field, size of the equatorial dead zone increases and the mid-latitude wind zone with open field lines decreases, which results in an inhibition of  atmospheric escape.} Increasing equatorial dead zones with increasing planetary field strength dominate the optically thick area that gives rise to the transit signal and as a result with increasing planetary magnetic field, transit depth increases. These results are corroborated by the study of \citet{Owen2014}. Their findings were more quantitative in the sense that they put a lower limit on the magnetic field of planets. If the planetary magnetic field $B_{\rm p} \geq 0.3G$, the flow would be controlled magnetically even for the highest incident UV flux $\sim$ 10$^6$ erg~s$^{-1}$. The mass-loss rate is in their simulations suppressed approximately by an order of magnitude in comparison to non-magnetic case \citep{Owen2014}.    

The first 3D MHD simulation of the interaction of magnetized stellar environments with magnetized hot Jupiter atmosphere has been performed by \citet{Cohen2011}. They have studied star-planet interaction and magnetic reconnection part mainly including the possibility of a generation of hot spots on the stellar surface. The most rigorous magnetic interaction of stars with hot Jupiter atmosphere, which includes stellar and planetary outflows was carried out by \citet{Matsakos2015} using the PLUTO code. They have classified the star-planet interaction into four types. In type-I interaction, if the planetary outflow is weak,  the stellar plasma is intercepted by the planetary magnetic field. The relative motion between the planet and the stellar wind gas is typically large enough to form a bow shock upstream of the planet. The shocked stellar plasma sweeps back the planetary field and diverts the planetary outflow in the downstream direction (see top left panel of figure~13 of \citet{Matsakos2015}). In type-II and type-III interaction, stellar wind collides with the strong planetary outflow that opens up the magnetosphere. The shocked planetary materials are dragged backwards by the stellar outflow forming a wide tail. The difference between type-II and type-III interaction is that for type-II, the interaction happens within the Roche lobe and shocked planetary materials remain confined but for type-III, it happens outside of the Roche lobe resulting in the accretion of shocked planetary plasma towards the star. In the type-IV interaction, stellar plasma is intercepted by the planetary magnetic field outside of the Roche radius causing Roche-lobe overflow. As the gas in their star-planet interaction simulation is fully ionized, they concluded with the note that the neutral hydrogen (H I) or magnesium line (Mg I) would not be detected from this system.   

\begin{figure}[htbp!]
\centering
\includegraphics[width=1.0\textwidth]{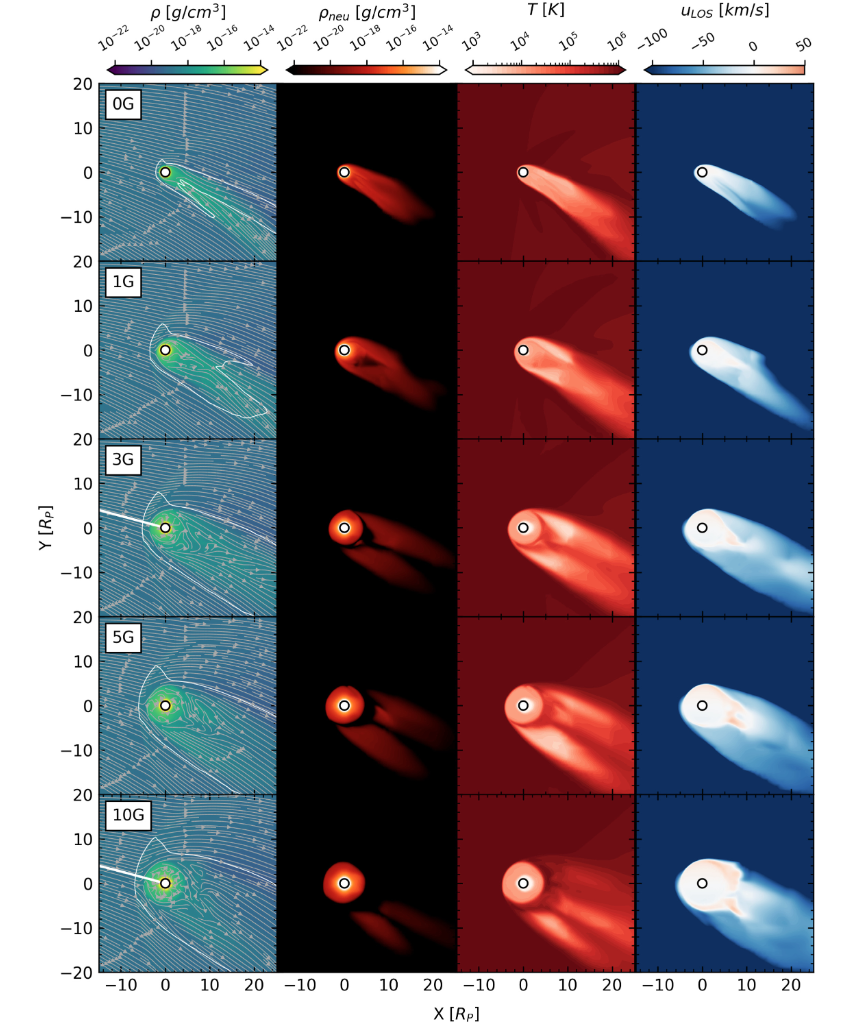}
\caption{Behaviour of the planetary atmosphere during the interaction with magnetized stellar wind. The star is kept outside of the simulation grid and located at negative x. The stellar wind is entering from the left side into the simulation grid. The planet is kept at the center of the grid and marked by a circle. Each row shows different planetary magnetic dipole field strengths. The first column shows the total density profile. The grey contours show the velocity streamlines and the magnetosonic surface is shown in the white contour. The second, third, and fourth columns show neutral hydrogen density, temperature, and line-of-sight velocity of the planetary atmosphere. The figure is taken from \citet{Carolan2021}}\label{fig:magnetic_filed_escape}
\end{figure}

Recently \citet{Carolan2021} studied the effect of magnetic field on atmospheric escape using 3D radiative MHD simulations. This is the model that solves the same set of equations as given in equation~\ref{eq:cont_3D} - \ref{eq:energy_3D} with magnetic field added (see details in \citet{Carolan2021}). The planetary magnetosphere is assumed to be dipolar and the planetary magnetic field strength is varied from 0 to 10 G. The stellar wind magnetic field and geometry are kept the same for all cases of simulation. Figure~\ref{fig:magnetic_filed_escape} shows the atmospheric structure of a hot Jupiter with different planetary dipolar magnetic field strengths. The first, second, third, fourth, and fifth rows show the cases with a planetary magnetic field strength of 0, 1, 3, 5, and 10 Gauss respectively. The column from left to right consecutively shows the total density, neutral hydrogen density, temperature, and line-of-site total velocity of the planetary materials. Unlike 2D MHD models \citet{Trammell2014, Owen2014}, the mass-loss rate does not change much with the increasing planetary magnetic field. They found a small increase in escape rate with the planetary magnetic field. However, the dynamics of the planetary atmosphere changes with increasing field strength. With increasing field strength, the size of the equatorial dead zone increases and produces a double-trail structure. This double-tail structure in atmospheric escape simulation is the first prediction from this model and requires intensive observational studies to support this. 

The Ly-$\alpha$ transit calculation in their model shows an enhancement in line-center absorption with increasing planetary magnetic field strength, as more planetary materials are trapped in the increasing equatorial dead zone. The absorption in the blue wing due to high-velocity neutral was also found to be increasing with planetary field strength.

Some studies \citep[e.g.,][]{Cohen2015, Dong2017} on the effect of magnetic field on the terrestrial exoplanets are also performed. \citet{Cohen2015} performed a simulation of the interaction of magnetized stellar wind on the unmagnetized Venus-like planets and concluded that the stellar wind can penetrate very deep into the atmosphere and enhance the mass-loss rate. For proxima-centauri b, \citet{Dong2017} reported that the mass-loss rates are lowered if the planet has an intrinsic dipole magnetic field. A detailed study of terrestrial exoplanets for the magnetic interaction of stellar wind with the planetary magnetosphere is needed. Also, special care is needed to interpret those results from transit observations which is missing from the present studies.

\subsection{Impact of flares and Coronal Mass Ejections(CMEs) of host star on exoplanetary atmospheric escape}\label{sec:flare_cme}
The flares and CMEs when occurring on the host star change the radiation, and plasma density of the outflowing wind from the host star. The modified radiation during flares from the host star changes the photochemistry and ionization level of the upper atmosphere of the planet. On the other hand, the CME enhances the dynamic pressure, charge exchanges, and collisional ionization on the planetary upper atmosphere. As a result, both of the events increase the planetary mass-loss rates. Hence the study of the effect of flares and CMEs on the planetary mass loss is very important and has been carried out by several groups \citep{Segura2010, Venot2016, Kay2016, Chadney2017, Cherenkov2017, Bisikalo2018, Hazra2020, Louca2022, Hazra2022}. The changes in the planetary mass-loss rate during the flares and CME would also change the observational signature temporarily. The Ly-$\alpha$ transit observations of HD189733b by \citet{Lecavelier12} using STIS, HST in two epochs show an enhancement of transit depths in the second epoch compared to the first epoch. The first transit observation that was made in April 2010 showed a negligible transit depth. The second transit event in September 2011 showed an enhanced transit absorption depth of $14.4 \pm 3.6 \%$ between velocities of -230 to -140 km~s$^{-1}$. At the second epoch, 8 hours before the transit observation, an X-ray flare from the star was also observed with Swift/XRT. These combined observations support the idea that the individual flare or flare-associated CME must have caused an enhanced mass loss from the upper atmosphere of the planet HD189733b \citep{Guo2016, Chadney2017, Hazra2020, Hazra2022}.      

For gaseous giant planets, \citet{Hazra2020} analyzed the effect of flares on the upper atmosphere using a 1D escape model and found an enhancement of mass loss due to flaring events. The transit signatures in both Ly-$\alpha$ and H-$\alpha$ have been enhanced due to the flare on the host star. As the 1D model can not take into the interaction of stellar wind, the asymmetric high-velocity blue and red wings could not be modeled. 

\begin{figure}[htbp!]
\centering
\includegraphics[width=1.0\textwidth]{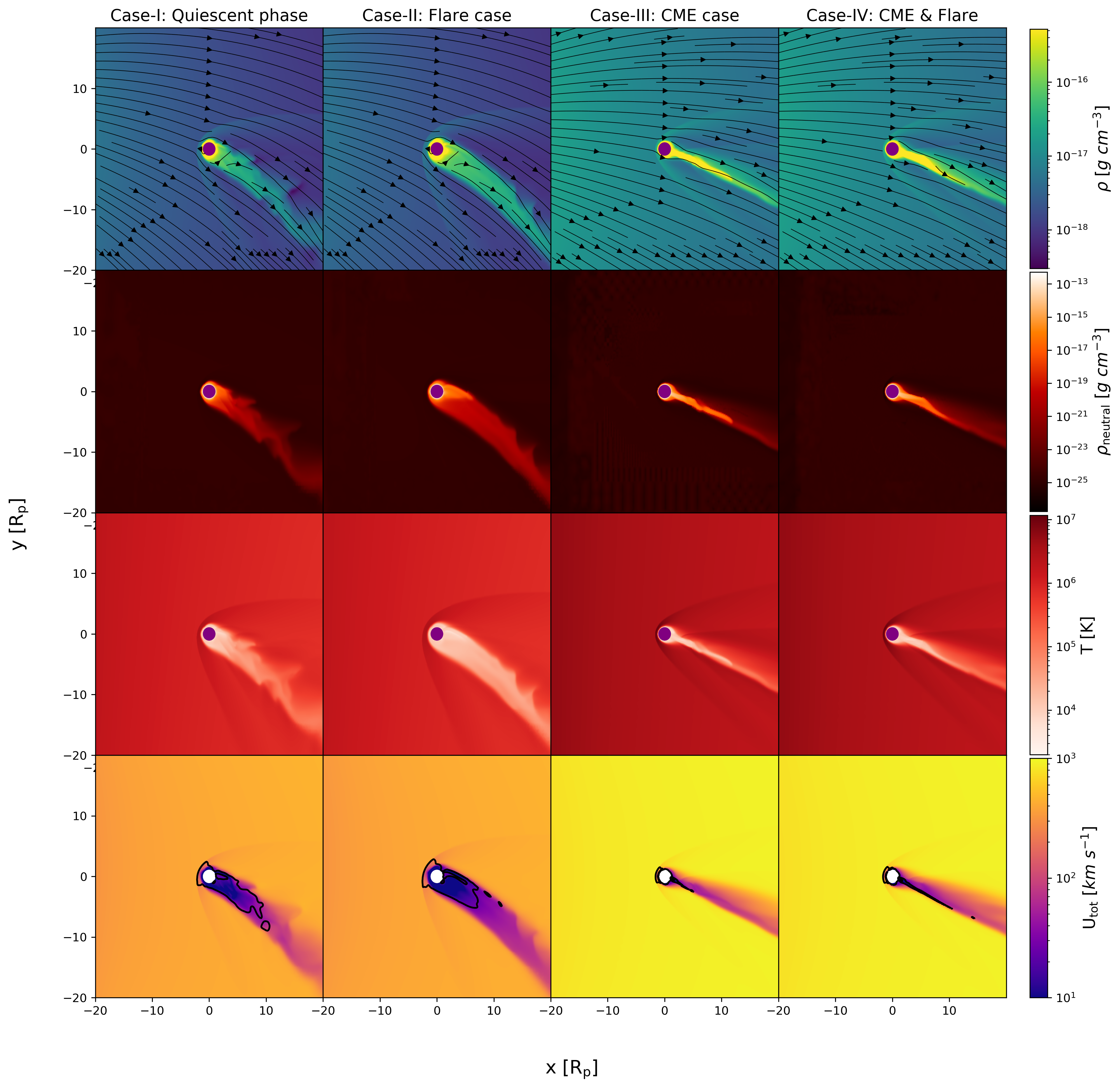}
\caption{The behavior of the planetary atmosphere for the four cases is shown here. The first column shows the total density (first row), neutral density (second row), temperature (third row), and total velocity (fourth row) for the quiescent stellar wind case. The second column, third column, and fourth column show the same quantity in each row for case-II, case-III, and case-IV respectively. All the plots are shown in the orbital planes (z = 0).}\label{fig:flare_cme}
\end{figure}

A more rigorous calculation of the effect of flare and CME on the planetary atmosphere of a hot Jupiter has been performed by \citet{Hazra2022} using a 3D radiation-hydrodynamics model. They investigated four cases: first - stellar wind in the quiescent phase of the star, second - a flare, third - a CME, and fourth - a flare that is followed by a CME. The dynamics of the planetary atmosphere for these four cases are shown in figure~\ref{fig:flare_cme}. Each row shows different planetary properties. The total density, neutral density, temperature, and total velocity are shown from the first to the fourth row, respectively. Each column shows each of the considered four cases. The behavior of planetary materials is different for the four cases. The distinct difference arises when we consider the CME case. Comparison of the behavior of different atmosphere properties for case-1 and case-III clearly shows the difference in the cometary tail where materials are swept out by stellar wind and CME plasma respectively. For the flare case (case-II), the day-side sonic surface resides a bit higher than the quiescent stellar wind case (case-III) because of the higher rate of planetary outflow. The mass-loss rates for the four cases differ quite significantly. The flare case alone increases the mass-loss rate by 25\% compared to the quiescent phase, while the CME leads to an increment by a factor of 4. 

\begin{figure}[htbp!]
\centering
\includegraphics[width=0.95\textwidth]{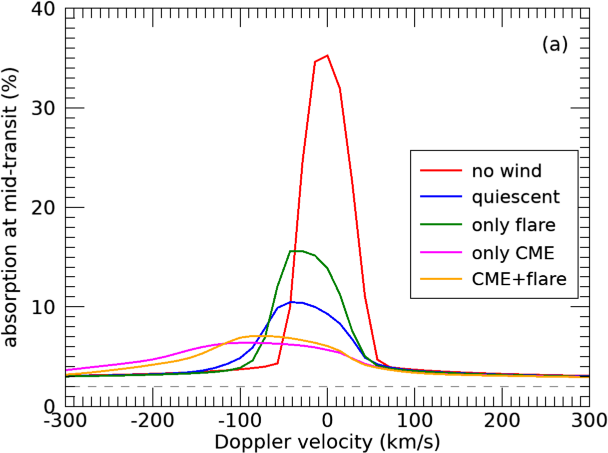}
\caption{ Synthetic Ly-$\alpha$ spectra calculated from planetary transit for four cases mentioned above. The blue, green, magenta, and yellow lines show the transit spectra for the quiescent case, flare case, CME case, and CME+flare case. The red line shows the transit spectra due to radiation-driven planetary outflow alone without any interaction with stellar wind}\label{fig:flare_transit}
\end{figure}

The synthetic Ly-$\alpha$ simulated for the four cases mentioned above also show different transit absorption depths, which is shown in figure~\ref{fig:flare_transit}. The flare case, CME case, and CME + flare case are shown using green, magenta, and yellow lines respectively. The flare case increases the transit depth in the line center but the CME case increases the transit depth in the blue-wing (-230 to -140 km~s$^{-1}$). Hence \citet{Hazra2022} concluded that the observed enhanced transit depth in the second epoch of the transit measurement \citep{Lecavelier12} is more likely due to a CME affecting the planetary atmosphere not due to the flare alone as argued before. 

\citet{Louca2022} has recently studied the effect of stellar flares on the chemistry of gas giants to terrestrial exoplanets. Significant changes are reported in the abundance of some relevant species such as H, OH, and CH$_4$. They reported the abundance changes up to three orders of magnitude in comparison to pre-flare abundances. \citet{Konings2022} showed that the flares can affect the atmosphere of gas giants up to several days after the flare event during which the transmission spectra are altered by several hundred ppm. {However, recently \citet{Gillet2025} studied the effect of flares on gas giants considering a realistic temporal evolution of the XUV flux of a flare and found no effect of flares on mass-loss rate and Ly-$\alpha$ signature. Hence, to understand effect of flares on exoplanet atmospheres, more thorough studies are needed.}

Most of the previous efforts to study the impact of CMEs on atmospheric escape were done using hydrodynamical models \citep[e.g.,][]{Kay2016, Cherenkov2017, Hazra2022}. Recently \citet{Hazra2024} studied the magnetic interaction of CME plasma with the atmosphere of a hot Jupiter with dipolar magnetosphere using time-dependent radiative MHD simulations. They have considered three configurations of embedded CME magnetic field - (a) northward $B_z$ component, (b) southward $B_z$ component, and (c)radial component. They found that CMEs with both northward and southward $B_z$ are most effective in removing planetary material and hence escape rate. The planetary magnetosphere goes through three distinct changes while interacting with the CME. Interestingly, the maximum Ly-$\alpha$ transit absorption depth is found when the CME has a radial embedded field and less planetary escape rate.   

\subsection{A note on computing synthetic transit spectra from atmospheric escape models}
To probe atmospheric escape from exoplanet observationally and constrain theoretical models, transit spectroscopy is the only reliable methodology available to us. As I explain in section~ \ref{sec:numerical_model}, most of the numerical models (both 1D and multidimensional) compare their result every time with the transit spectra of Ly-$\alpha$, as atmospheric escape have been studied most in the hydrogen Ly-$\alpha$ line. 
{There are also some theoretical models \citep[e.g.,][]{Hazra2020, Mitani2022, Yan2022, Yan2024} that compare their simulated results with H-$\alpha$ transit observations.  \citet{Hazra2020} synthetically calculated H-$\alpha$ line for different XUV radiation during a magnetic cycle of host star and estimated a variation of 1.9 m\AA\ in equivalent width over the cycle. They also found that H-$\alpha$ increases during a flare case.\citet{Yan2022} and \citet{Yan2024} modelled WASP-52b and HAT-P-32b using 1D hydrodynamic model to compute the H-$\alpha$ transmission spectra combined with non local thermodynamic and Monte Carlo simulation. By comparing H-$\alpha$ observation with simulated H-$\alpha$ spectra, they show that it is possible to constrain the XUV radiation and metallicity of the atmosphere. Using 3D model \citet{Mitani2022} computed H-$\alpha$ line due to the stellar wind interaction with the upper atmosphere of exoplanet and found a weak effect of stellar wind on H-$\alpha$ line.}
After the first detection of helium escaping a planet \citep{Spake2018}, constraining theoretical models using helium lines becomes relevant as the helium lines would help us to constrain the launching region of the planetary outflow. {There are many 1D hydrodynamic escape models \citep[e.g.,][]{DosSantos2022, Lampon2020, Zhang2023,Lampon2023, Rosener2025} have been developed in the last few years which compute helium signals.  However, 3D numerical models of helium escape modeling have been limited to date and we need more self-consistent 3D models.} Another interesting constraint on the theoretical model might come from the modeling of transit spectra using metal lines, which also has not been done extensively. {Although in the last few years, Ly-$\alpha$ transit observations gave us a lot of constraint on the mass-loss rate from exoplanets, with the discovery of helium and the possibility of detecting it from the ground based telescopes, helium triplet line could be used to probe the atmospheric escape widely. Actually, the most suitable things would be simultaneous modeling of hydrogen and helium lines and compare them with observations. Because of the contamination of Ly-$\alpha$ by interstellar medium and geo-emission, observing H-$\alpha$ and helium triplet line should be used as a complementary method for understanding evaporation from exoplanets \citep{Czesla2022, Yan2022, Yan2024}.}


An important point to note while interpreting observed Ly-$\alpha$ absorption as the actual mass-loss rate is that Ly-$\alpha$ transit occurs due to the absorption of stellar Ly-$\alpha$ line by neutral hydrogen only. Hence Ly-$\alpha$ should give the mass-loss rate of neutral hydrogen. It does not give any information about the ionized hydrogen (proton) loss. Recent non-detection of Ly-$\alpha$ in many highly irradiated close-in exoplanets (HD97658b \citep{Bourrier2017}, $\pi$ Men C \citep{GarciaMunoz2020}, young planet K2-25b \citep{Rockcliffe2021}, HD63433b \citep[the third planet HD63433c further away than the second planet HD63433b shows significant Ly-$\alpha$ absorption]{Zhang2022a}) needs further attention and investigation as to why Ly-$\alpha$ will not always be an indicator for mass loss. \citet{Owen2023} analysed the fundamentals of Ly-$\alpha$ transit to interpret the observation. They found that for Ly-$\alpha$ to be detected, the planetary material must be sufficiently radially accelerated before it gets too ionized. The too much ionization of the cometary tail might be the reason for several non-detections of Ly-$\alpha$ in the close-in planets. Along with the transit depth, the transit duration is very important as it encodes information about the progressive ionization of the cometary tail. Another alternative explanation of the non-detection of Ly-$\alpha$ is stellar wind confinement \citep{Vidotto2020, Carolan2021a}. The stellar wind of the host star is so strong that it confines the planetary outflow to an undetected limit. These studies \citep{Owen2023, Carolan2021a} make it clear that the non-detection of Ly-$\alpha$ does not mean that the planet is not going through atmospheric escape, nor is the planet hydrogen-dominated, rather it needs careful interpretation. {Much more constraints could be imposed on the evaporating atmosphere if we observe the hydrogen and helium line simultaneously \citep{Czesla2022, Bello-Arufe2023}. Not only simultaneous observation will give us confirmation that the atmosphere is indeed evaporating but also it will help us constraining the H/He ratio in the atmosphere and the XUV radiation from the host star. \citet{Yan2022} and \citet{Yan2024} studied atmosphere of WASP-52b and HAT-P-32b in both H-$\alpha$ and helium triplet simultaneously. For WASP-52b, to match with the observation of both H-$\alpha$ and helium triplet, a high XUV radiation is needed with a low
X-ray fraction in XUV radiation or less XUV radiation with high X-ray fraction with H/He ratio = 98/2 \citep{Yan2022}. For HAT-P-32b, they found a solar-like metallicity in the upper atmosphere \citep[see][for details]{Yan2024}. Nevertheless, in the near future, simultaneous modeling of hydrogen and helium lines will pave the way for understanding the evaporating atmosphere and its compositions.}

\section{Importance of atmospheric escape in the evolution of exoplanets}\label{sec:evolution}
The atmospheric escape plays the most important role in atmospheric erosion which sometimes leads to the complete loss of the planetary atmosphere. Hence, to understand planetary atmospheric evolution, it is very important to understand the atmospheric escape mechanism properly. Some earlier studies \citep{Khodachenko2007, Lammer2007} showed that the CME-driven atmospheric escape can strip off the whole atmosphere of unmagnetized exoplanets during a 1-Gyr period. With the recent discovery of radius gap \citep{Fulton2017} and Neptunian desert \citep{Mazeh2016} in the observed exoplanet demographics, studies of atmospheric evolution become very relevant, where evolution is invoked as one of the major reasons for these observed gaps in exoplanet demographics.  

Recent studies have shown that the observed radius valley can be reproduced by incorporating atmospheric mass loss in the evolution by either photoevaporation or core-powered mass loss \citep{Rogers2021}. Hence direct evidence of atmospheric evolution is verified by reproducing the radius valley as observed. Similarly, for the Neptune-sized planet, atmospheric mass loss over time completely removed the atmosphere of the Neptune-sized planet and gave rise to the observed Neptunian desert \citep{Ionov2018, Vissapragada2022}. However, this atmospheric evolution estimation due to mass-loss by photoevaporation or core-powered has assumed a parametric mass-loss formula, which may not be accurate as it is not easy to discern which types (e.g., photon-limited, energy-limited, or radiation/recombination-limited) of atmospheric escape is going on in the planet. Also, as we describe in section~\ref{sec:3Dmodel}, the stellar wind, stellar magnetic field, stellar flares, and CMEs play a very important role in mediating atmospheric mass-loss rate. To properly calculate atmospheric mass loss and its effect on overall planetary atmosphere evolution, we need to incorporate the full stellar environment and its interaction with the planetary atmosphere.

\section{Future perspective and outlook}\label{sec:conclusion}
In this review, I summarise the atmospheric escape processes from exoplanets including the methodology of their observation and theoretical modeling to date. The atmospheric escape from exoplanets has been driven largely by stellar environments - such as stellar radiation, stellar wind, CMEs, and flares \citep{Murray-Clay2009, Carroll-Nellenback2017, Hazra2022}. Sometimes Core-powered energy that the planet acquires during the accretion phase of planet formation can also lead the atmospheric escape \citep{Gupta2020}, but direct observational evidence is far from constraining this mechanism. 

One of the major driving forces of atmospheric escape from planets is stellar radiation (XUV radiation, infrared). The observational measurements of this important quantity are still in the very preliminary stage. We have measurements of the X-ray part but the extreme ultraviolet part of the stellar spectra is still observationally ill-constrained. The earlier efforts were to reconstruct the EUV part of the stellar spectra using empirical relations from X-ray \citep{Chadney2015} and stellar Ly$\alpha$ \citep{Sanz-Forcada11}. Recently there have been some efforts- MUSCLES Treasury Survey \citep{France2016, Youngblood2016} and MEGA MUSCLES survey \citep{Froning2019, France2020} to make a spectral database of many planet-hosting stars. In the MUSCLE survey, SEDs with a spectral range of 5 \AA\ to 5.5 $\mu$m of 11 nearby planet-hosting stars (M-dwarfs and K-dwarfs) in the solar neighborhood (d $\leq$ 20 pc) have been carried out. The MUSCLE survey includes observations in visible wavelength from Hubble and ground-based observatory, X-rays from Chandra/XMM-Newton/Swift. Other wavelengths in infrared and astrophysically inaccessible wavelengths, EUV and Ly-$\alpha$ are reconstructed using empirical relations and stellar model spectra. Recently Mega-MUSCLE survey \citep{France2020} added another 13 stars in the MUSCLE database. \citet{Behr2023} extended the database by adding SEDs of 11 JWST targets. Although MUSCLE database{\footnote{https://archive.stsci.edu/prepds/muscles/}} provides SEDs of some important planet-hosting stars, we need more observations of SEDs from parent stars in the near future. Also, the reconstruction of the observationally inaccessible part of the spectra needs a better physics-constrained model instead of mere empirical scaling relations \citep[e.g.,][]{Sanz-Forcada11, Linsky2014}. 

Immediately after the stellar radiation, the most important quantity that affects the atmospheric escape (non-thermal loss) is the stellar wind. Depending upon the stellar wind strength, the mass-loss rate of the escaping atmosphere is decided (see section~\ref{sec:sw} for details). The strength of the stellar wind is approximated by the mass-loss rate from stars and observational quantification of this quantity would be very important for atmospheric escape calculation from exoplanets. By measuring astrospheric absorption signature produced by the interaction between the stellar winds and the interstellar medium, \citet{Wood2021} estimated the mass-loss rate of M-dwarf stellar winds. The soft X-ray emission (produced by charge exchange between heavy ions of the stellar wind and cold neutrals of interstellar medium) from the stellar astrosphere is also another good detector of stellar wind mass-loss rate \citep{Kislyakova2024}. A detailed study of observational stellar wind with theoretical modeling is needed to constrain stellar wind strength, which will help to probe atmospheric escape from the planets appropriately.    

Many numerical simulations \citep{Lammer2007, Khodachenko2007, Kay2016, Bisikalo2018,  Hazra2022, Odert2020, Hazra2024} showed that coronal mass ejections and flares can significantly alter (sometimes one order of magnitude) the escape rate from the planetary atmosphere. Observations of the physical properties of CMEs and their occurrence rate in planet-hosting stars would be the next important steps in understanding their effect on atmospheric escape. Recently \citet{Veronig2021} detected coronal dimming associated with flares on cool stars, which paves the way for comprehensive characterizations and detection of stellar CMEs. As flares and CMEs are directly affecting the atmospheric escape, their higher occurrence rates would lead to higher atmospheric mass loss from the exoplanets, which is crucial for atmospheric evolution. For the young host stars, flares occur more frequently than the stars which are of solar ages \citep{Audard2000, Maehara2015}. However, for the stellar CME, we have no observational information on the occurrence rate. For the Sun, the CME occurrence rate varies from 0.5 per day near solar minima to $\sim$ 6 near solar maxima \citep{Gopalswamy2003, Yashiro2004}. Some of the studies \citep{Drake2013, Odert2017} have shown that young and more active stars could have a higher rate of CME occurrence than the Sun but this needs further careful study.  

Along with constraining stellar environments from observations, we need detailed observations to understand atmospheric escape processes on the planetary side too. As we explained details in section~\ref{sec:observations}, the transit observations have given us atmospheric escape rates \citep[eg.,][]{Vidal-Madjar2003, Vidal-Madjar04, Fossati2010, Lecavelier12, Bourrier2016} from many exoplanets. However, the detailed atmospheric structure is yet to be probed from observations including the exobase location. Most of the observations were carried out using hydrogen Ly-$\alpha$ lines, which can not probe low-velocity escaping planetary materials. {On the other hand, helium triplet and H-$\alpha$ lines can probe the low-velocity launching area of escaping materials and have advantage of being observed from the ground based telescopes. 
Hence we need a simultaneous measurement of the escaping atmosphere by both helium and H-$\alpha$ transit lines in the near future for details understanding of the atmospheric escape. A combined Ly-$\alpha$ measurement along with H-$\alpha$ and helium line would be appropriate to constrain the  H/He ratio in the atmosphere and the stellar XUV radiation.}

Theoretical models for atmospheric escape are becoming mature with time. Atmospheric escape is an inherently 3D phenomenon and modeling them using 1D models would simplify many features, however, for the long-term planetary evolution studies, 1D models have been very helpful \citep{Kubyshkina2018, Johnstone2018, Allan2019}. In the 3D modeling, the inclusion of multispecies (e.g., Hydrogen, Helium, Oxygen, Carbon, etc) definitely would help for the interpretation of the observed transit spectra and underlying physics, which is an ongoing effort \citep{Dong2018, Hazra2022, Hazra2024}. However, in the presence of a magnetic field on the planet, different ionized species would behave differently with different velocities, and developing multi-fluid models would be the ultimate goal for a sophisticated atmospheric escape model.  {There are some efforts in this direction using 1D multifluid models \citep[e.g.,][]{Guo2019, Xing2023}. While studying water escape from terrestrial exoplanet using a 1D multifluid model, \citet{Guo2019} found that the critical XUV value might be increased to a factor of 2 in case we neglect the ion momentum transfer, as ions effectively transfer momentum between interacting species. This multifluid model also helps us explain the mass fractionalization of helium in the escaping atmosphere of hot Jupiter HD209458b \citep{Xing2023}}. Multidimensional multifluid models will help to understand charge exchanges \citep{Esquivel2019} by stellar wind/CME plasma with the planetary materials and pick up ions by stellar wind. For the modeling of atmospheric escape from rocky exoplanets, we may sometimes need to go beyond fluid codes and use fully kinetic or hybrid codes to model atmospheric escape and its interaction with stellar wind \citep[e.g.,][]{Jarvinen2018}.   

With the improved observations and modeling, we are progressing towards an understanding of atmospheric escape, but many questions remain due to the severe complexity involved in the atmospheric escape process. One of the main questions is related to radiative transfer modeling of incident stellar radiation. For the tidally locked exoplanets, the plane parallel approximation of radiation might not be a bad approximation but in the case of non-tidally locked planets, a better treatment of incident radiation is necessary. { \citet{Yan2022} for the first time performed transmission spectroscopy using a spherical stellar Ly-$\alpha$ radiation
source. They found that plane-parallel illumination models overestimate the Ly-$\alpha$ strength for the plane planetary atmosphere and underestimate for spherical planetary atmosphere in comparison to those for the spherically illuminated models.}

The coupling between the lower atmosphere and upper atmosphere should also be carefully considered as the abundance of various species in the upper atmosphere also depends on their values in the lower atmosphere. The role of the planetary magnetic field in atmospheric escape is very important and needs further study in the near future. Whether a planet with a magnetic field can protect an atmosphere from escaping or help the atmosphere in escaping - this conundrum needs to be tackled with details numerical modeling, which will include different values of planetary magnetic field and their different orientation of magnetosphere. is the role of the planetary magnetic field different for gaseous exoplanets to rocky exoplanets? In hindsight, a combination of observational studies along with self-consistent 3D radiative MHD or kinetic modeling of atmospheric escape is the key to solving many unresolved mysteries of atmospheric escape from exoplanets.  


\bmhead{Acknowledgements}
The author thanks two anonymous referees for their constructive comments that helped a lot to improve the manuscript. The author would also like to thank Prof. Mitsuru Kikuchi for his kind invitation to write this review. The author also acknowledges the IIT Kanpur initiation grant (IITK/PHY/2022386) for financial support. 

\section*{Declarations}
{Conflict of interest} The author declares no conflict of interest



\bibliography{my_ref, myref_exo2}

\end{document}